\documentclass[manuscript]{aastex}
\usepackage{natbib,longtable,hyperref}

\newcommand{\beq}{\begin{equation}}
\newcommand{\eeq}{\end{equation}}

\begin{document}

\title{WISE/NEOWISE Observations of the Hilda Population: Preliminary Results}

\shorttitle{NEOWISE Observations of the Hildas}
\shortauthors{Grav {\it et al.}}
\medskip

\author{T.~Grav} 
\affil{Department of Physics and Astronomy, Johns Hopkins University, Baltimore, MD21218, USA; tgrav@pha.jhu.edu}
\author{A.~K.~Mainzer, J.~Bauer\altaffilmark{1}, J.~Masiero} 
\affil{Jet Propulsion Laboratory, California Institute of Technology, Pasadena, CA 91109, USA}
\author{T.~Spahr}
\affil{Minor Planet Center, Harvard-Smithsonian Center for Astrophysics, Cambridge MA 02138, USA}
\author{R.~S.~McMillan}
\affil{Lunar and Planetary Laboratory, University of Arizona, Tucson, AZ 85721, USA}
\author{R.~Walker}
\affil{Monterey Institute for Research in Astronomy, Marina, CA 93933}
\author{R.~Cutri}
\affil{Infrared Processing and Analysis Center, California Institute of Technology, Pasadena, CA 91125, USA}
\author{E.~Wright} 
\affil{UCLA Astronomy, PO Box 91547, Los Angles, CA 90095, USA}
\author{P.~R.~Eisenhardt} 
\affil{Jet Propulsion Laboratory, California Institute of Technology, Pasadena, CA 91109, USA}
\author{E.~Blauvelt, E.~DeBaun, D.~Elsbury, T.~Gautier~IV, S.~Gomillion, E.~Hand}
\affil{Jet Propulsion Laboratory, California Institute of Technology, Pasadena, CA 91109, USA}
\author{A.~Wilkins\altaffilmark{2}}
\affil{Department of Astronomy, University of Maryland, College Park, MD 20742}

\altaffiltext{1}{Infrared Processing and Analysis Center, California Institute of Technology, Pasadena, CA 91125, USA}
\altaffiltext{2}{Jet Propulsion Laboratory, California Institute of Technology, Pasadena, CA 91109, USA}

\date{\rule{0mm}{0mm}}
\begin{abstract}
We present the preliminary analysis of 1023 known asteroids in the Hilda region of the Solar System observed by the NEOWISE component of the Wide-field Infrared Survey Explorer (WISE). The sizes of the Hildas observed range from $\sim 3 - 200$km. We find no size - albedo dependency as reported by other projects. The albedos of our sample are low, with a weighted mean value $p_V = 0.055\pm0.018$,  for all sizes sampled by the NEOWISE survey. We observed a significant fraction of the objects in the two known collisional families in the Hilda population. It is found that the Hilda collisional family is brighter, with weighted mean albedo of $p_V = 0.061\pm0.011$, than the general population and dominated by D-type asteroids, while the Schubart collisional family is darker, with weighted mean albedo of ($p_V = 0.039\pm0.013$). Using the reflected sunlight in the two shortest WISE bandpasses we are able to derive a method for taxonomic classification of $\sim 10\%$ of the Hildas detected in the NEOWISE survey. For the Hildas with diameter larger than 30km there are $67^{+7}_{-15}\%$ D-type asteroids and $26^{+17}_{-5}\%$ C-/P-type asteroids (with the majority of these being P-types).
\end{abstract}
\section{Introduction}

The Hildas are a population of asteroids in the 3:2 mean motion resonance with Jupiter, so that their orbital semi-major axes are at $\sim4.0$AU. It is believed to be populated by low-albedo C-, P- and D-type asteroids \citep{Gradie.1989a}. Due to the heliocentric distance of the Hildas, they are believed to have experienced less heating and are assumed to be of more pristine composition than objects in the main belt. \citet{Jones.1990a} found that the P- and D-types appear anhydrous and \citet{Luu.1994a} were unable to find any absorption bands in the infrared that were indicative of organics. More recently, however, near-infrared spectra of several D-type Jovian Trojans have been reported containing these bands \citep{Emery.2003a} and the hydration band near $3\mu$m has been reported only for a few inner main belt P- and D-types \citep{Rivkin.2002a,Kanno.2003a}. Both P- and D-types may, however, contain significant amounts of hydrosilicates without showing any detectable absorption bands if their surfaces are rich in opaque phases \citep{Cruikshank.2001a}. \citet{Carvano.2003a} pointed out that inner belt D-type objects often have concave spectral shapes and higher albedos compared to the outer belt D-types, suggesting that they may be compositionally different. Thus, at present the Hildas are assumed to be composed of a mixture of organics, anhydrous silicates, opaque material and ice \citep{Bell.1989a, Gaffey.1989a, Vilas.1994a}, but with only one D- and no P-type analogues among the meteorites found on Earth it is very difficult to accurately determine their compositions \citep{Hiroi.2001a}. 

The composition of outer belt asteroids by CCD spectroscopy has been studied by \citet{Vilas.1985a}, \citet{Dahlgren.1995a} and \citet{Dahlgren.1997a}. These authors found that the D-type objects make up $34\%$ of the numbered Hilda asteroids at that epoch, with the P- and C-types made up $28\%$ and $2\%$, respectively. They also found a spectral slope-asteroid size relation among the Hilda population, implying a size dependent surface composition where the P-types dominate at larger sizes. They suggested that the main size-dependent physical process acting on the Hildas are their mutual collisions, thus if D-types are more fragile than the P-types, this will favor disruptive collisions among the D-type precursors. \citet{Gil-Hutton.2008a} used the Sloan Digital Sky Survey (SSDS) sample of  122 Hilda asteroids to show that this size-taxonomy correlation appears to only be valid for $H<12$, i.e. the largest objects. 

Thermal observations of 23 Hildas were collected by InfraRed Astronomical Satellite \citep[IRAS;][]{Matson.1986a,Tedesco.1992a,Ryan.2010a}. \citet{Ryan.2011a} reported thermal observations of an additional 64 objects in the Hilda population collected with the {\it Spitzer Space Telescope}. They reported an apparent size and albedo dependency, with lower sizes yielding higher albedos. 

The population of minor planets in the first-order mean motion resonances with Jupiter, i.e. the Jovian Trojan, Hilda and Thule populations, are possible footprints of the orbital evolution of the giant planets. Stability or instability of these populations is directly related to the orbital configuration of the giant planets, and they also provide constraints and clues to the nature and amount of migration by Jupiter and other details of its dynamical behavior \citep{Broz.2008a}. The current configuration of the giant planets is the result of some dynamical evolution in the early solar system, and recently the so-called Nice model has gained significant traction in describing this evolution \citep{Gomes.2005a,Morbidelli.2005a,Tsiganis.2005a}. This model puts Jupiter and Saturn interior to their mutual 1:2 mean motion resonance \citep{Morbidelli.2007a} with the event of crossing this resonance having major influence on not only the final configuration of the planets, but also strongly affecting the distribution of minor planets. Studies by \citet{Morbidelli.2005a} and \citet{Broz.2008a} show that such a scenario would destabilize the Jovian Trojan and Hilda populations, then repopulating them later during the same phase of the dynamical evolution. Determination of the physical properties of the Hilda population (such as their numbers, sizes, albedos, and orbital distribution) is thus important as it allows us to compare them to that of the Main Belt Asteroids and Jovian Trojan populations. This could reveal clues as to whether the population is of primordial origin or was inserted into the resonance during the later stages of planet migration. In particular the difference or similarities between the Hilda and Jovian Trojans populations puts constraints on the origin of the bodies needed to repopulate the two resonances after the migration of Jupiter. 

In this paper we will try to answer the following questions: 1) What is the albedo distribution of the Hildas and is there a size-albedo relation as reported by \citet{Ryan.2011a}? 2) What is the size-frequence distribution? 3) What is the relative fraction of C-, P- and D-type asteroids in the Hilda population? In this paper we discuss the observations in Section \ref{sec:obs} and select our Hilda sample in Section \ref{sec:objsel}. The thermal model is described in Section \ref{sec:thermal} and the results are discussed in Section \ref{sec:res}. 

\section{Observations}
\label{sec:obs}

WISE is a National Aeronautics and Space Administration (NASA) medium-class Explorer mission designed to survey the entire sky in four infrared wavelengths, $3.4$, $4.6$, $12$ and $22$ $\mu$m \citep[denoted W1, W2, W3, W4 respectively;][]{Wright.2010a, Mainzer.2005a}. The survey collected over 2 million observations of more than 157,000 asteroids, including Near-Earth Objects, Main-Belt Asteroids, comets, Hildas, Jovian Trojans, Centaurs and scattered disk objects \citep{Mainzer.2011a}. With this sample, WISE has collected infrared measurements of nearly two orders of magnitude more asteroids than its predecessor, IRAS \citep{Matson.1986a,Tedesco.1992a,Tedesco.2002a}. The survey started on 2010 January 14 and exhausted its secondary tank cryogen on 2010 August 5. Exhaustion of the primary tank cryogen  occurred on 2010 October 1, but the survey was continued until 2011 February 1, as the NEOWISE Post-Cryogenic Mission, using only bands W1 and W2. 


The WISE observations of the Hildas were retrieved by querying the Minor Planet Center (MPC) observational files for all instances of individual WISE detections of the desired objects that were reported using the WISE Moving Object Processing System \citep[WMOPS;][]{Mainzer.2011a}. The WISE survey cadence resulted in most minor planets receiving on average 10-12 observations over $\sim 36$ hours \citep{Wright.2010a,Mainzer.2011a}. The resulting set of position/time pairs were used as the basis for a query of WISE source detections in individual exposures (known as Level 1b images) using the Infrared Science Archive (IRSA).  To ensure that only observations of the moving objects were returned from the query, a search radius of $0.3"$ from the observations in the MPC observation file was used. Since WISE collected a single exposure every 11 seconds, the modified Julian date was also required to be within 4 seconds of the time specified by the MPC. Only observations with 0 and p in the artifact identification \textsc{cc\_flag} were used, where 0 indicates no evidence of known artifacts were found at the position and p indicates that an artifact may be present. We have found that observations with \textsc{cc\_flags} of p are generally non-distinguishable from the non-flagged observations, indicating that the First-Pass version of the WISE data processing pipeline is very conservative in its artifact identification. Adding the observations flagged with p recovers about $20\%$ more observations.  Some of the Hildas have W3 magnitudes brighter than 4, at which point the detector approached experimentally-derived saturation limits. A linear correction was performed to account for the inaccuracy in the point-spread-function of these slightly saturated sources, and the W3 magnitude error was set to 0.2 magnitudes. 

In order to avoid having low-level noise detections and/or cosmic rays contaminating our thermal model fits we require that each object have at least three uncontaminated observations in a band. Any band that did not have at least $40\%$ of the observations of the band with the most numerous detections (in general W3 or W4 for the Hildas) was discarded even if it has 3 or more detections. WMOPS was designed to reject inertially fixed objects such as stars and galaxies in bands W3 and W4, but with stars having approximately 100 times higher density in bands W1 and W2, it is more likely that asteroid detections in these bands are confused with inertial sources. We remove such confused asteroid detections by cross-correlating the asteroid detections with sources in the WISE atlas and daily co-added catalogs from IRSA. Objects within $6.5"$ (equivalent to the WISE beam size at bands W1, W2 and W3) of the asteroid position appearing in the co-added sources at twice and in more than $30\%$ of the total number of coverages of a given area of sky were considered to be inertially fixed sources contaminating the asteroid photometry, and these observations were removed from the thermal fitting. 

\section{Object Selection}
\label{sec:objsel}

\begin{figure}[t]
\begin{center}
\includegraphics[width=12cm]{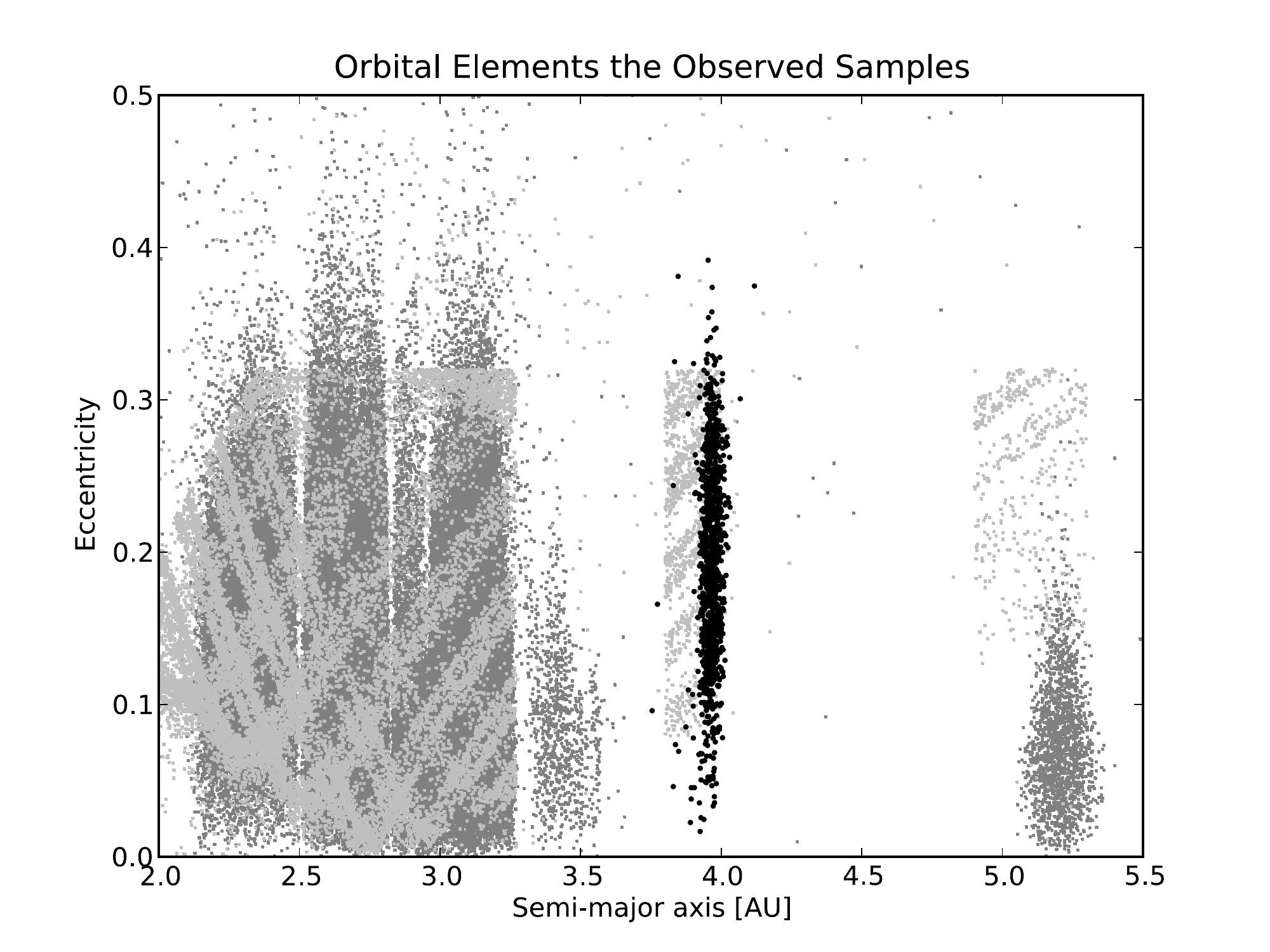}
\caption{The semi-major axes and eccentricities of the objects detected by NEOWISE. The black shows the Hilda objects with well established orbits that constitute the long-arc Hilda (LAH) sample. The dark gray gives the objects with observational arc lengths greater than 18 days, indicating that they were either already known at the time of WISE observation or subsequently had optical follow-up. The light grey indicates objects with observational arc lengths less than 18 days, and are generally WISE discoveries with no optical follow-up. The pattern seen in these short arc objects is an artifact of the procedures used by the MPC in deriving preliminary orbits for objects with such short arcs and is not a real property of the objects' orbits. }
\label{fig:aecc}
\end{center}
\end{figure}

\begin{figure}[t]
\begin{center}
\includegraphics[width=12cm]{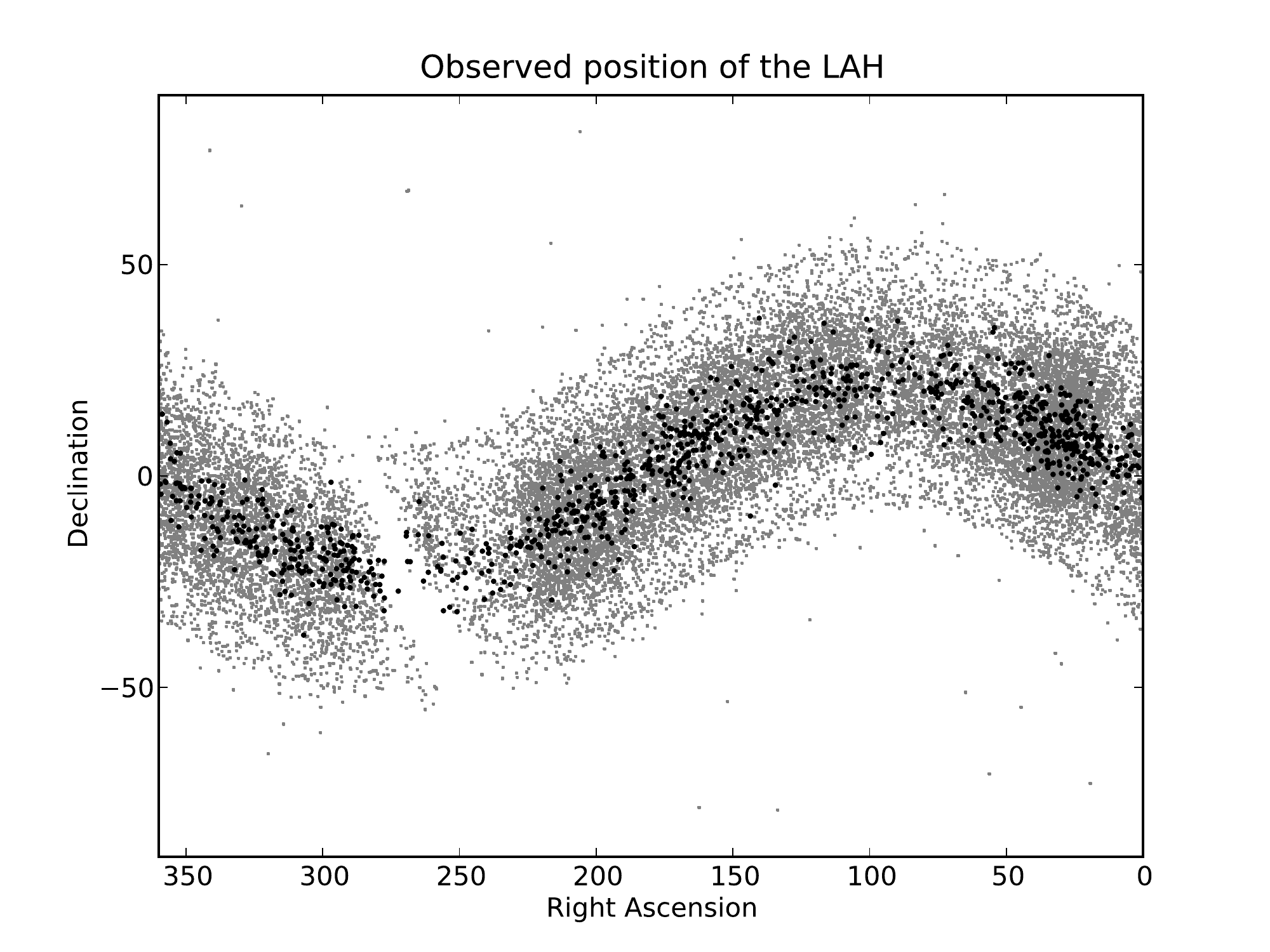}
\caption{The observed sky-plane position of the LAH (in black) compared to the short arc length sample (in gray).  }
\label{fig:radec}
\end{center}
\end{figure}

In this paper we will only consider the objects that have well determined orbits that securely define them as Hildas. The Hildas are in the 3:2 mean motion resonance with Jupiter, which lies at $a \sim 3.9$ (see Figure \ref{fig:aecc}). We define the Hildas in the most general sense, allowing their semi-major axis to be in the range $3.7 - 4.2$AU, with an eccentricity less than $0.4$ and an inclination less than $30^\circ$. To make sure that the orbits are generally secure we also require the observed arc length to be at least 18 days, which is longer than that needed for orbital determination to be able to differentiate between Main Belt Asteroids, Hildas and Jovian Trojans. There are 1028 objects in the dataset of objects observed by NEOWISE during the fully cryogenic part of the survey that satisfy these criteria (see Figure \ref{fig:radec}), and we label this sample the long arc Hildas (LAH). Of these, 923 objects were associated with previously known objects, while 105 objects were new discoveries that have subsequently been linked to incidental astronomy in the MPC one-night database or have received optical follow-up after the object was reported to the MPC.

We note that using these slightly relaxed criteria means that a handful of objects that are in Hilda-like orbits, that may not actually be in the 3:2 mean motion resonance, have been included in the LAH sample. Accurate long-term orbital integration of the Hildas to weed out these handful of objects, making up at most $1-2\%$ of the full sample, is beyond the scope of this paper, and the low number of objects in question are to few too significantly influence the results presented. 

\begin{figure}[t]
\begin{center}
\includegraphics[width=12cm]{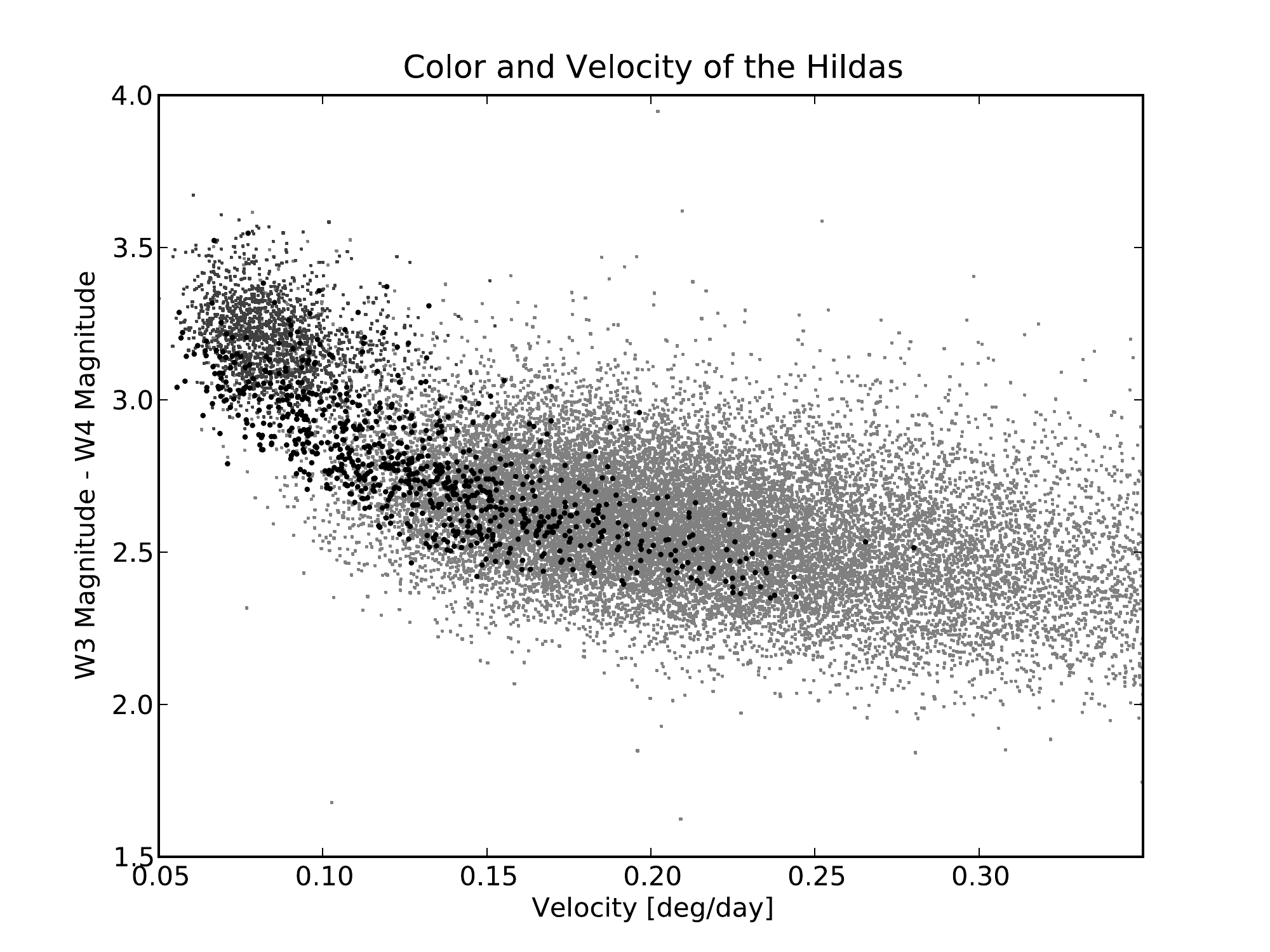}
\caption{The thermal color and observed sky-plane velocity of the LAH (in black) compared to the short observational arc sample (in gray).  }
\label{fig:w34vel}
\end{center}
\end{figure}

There is significant overlap in the observed WISE magnitudes between the Hildas and MBAs (see Figure \ref{fig:w34vel}). While in \citet{Grav.2011b} we were able to define a sample of candidate Jovian Trojans, the significant color overlap with the MBAs makes this unobtainable for the Hildas. Thus in this paper, we do not attempt to disentangle the objects in these populations with short observational arcs (less than 18 days). It is expected that current large sky surveys like Catalina Sky Survey, Lincoln Near-Earth Asteroid Research \citep[LINEAR;][]{Stokes.2000a} and the Panoramic Survey Telescope and Rapid Response System \citep[Pan-STARRS;][]{Wainscoat.2010a} will provide optical follow-up of a significant fraction of these objects in the next few years. At that point we will update the preliminary results presented in this work.

\section{Preliminary Thermal Modeling}
\label{sec:thermal}
Preliminary thermal models for each of the Hildas detected by WMOPS during the cryogenic portion of the survey and using the First-Pass Data Processing Pipeline \citep[version 3.5;][]{WISEexsup} described above have been computed (these models will be recomputed when the final data processing is completed sometime during the fall of 2011). As described in \citet{Mainzer.2011b}, the spherical near-Earth asteroid thermal model \citep[NEATM;][]{Harris.1998a} was used. The NEATM introduced the so-called beaming parameter $\eta$ to account for cases intermediate between zero thermal inertia \citep[the standard thermal model, or STM;][]{Lebofsky.1978a} and infinite thermal inertia \citep[the fast rotating model, of FRM;][]{Veeder.1989a,Lebofsky.1989a}. In the STM, $\eta$ is set to 0.756 to match the occultation diameters of Ceres and Pallas, while in the FRM, $\eta$ is equal to $\pi$. In the NEATM $\eta$ is a free parameter that can be fitted if two or more thermal bands are available, or using a single thermal band if {\it a priori} information of diameter and albedo is available from space craft or occultation observations. 

For each object a spherical surface was approximated using a set of triangular facets \citep[c.f. ][]{Kaasalainen.2004a}. While some Hildas are non-spherical, the WISE observations generally consist of 8-10 observations uniformly distributed over $\sim 36$h for each object, so any rotational variation is generally average out and the model yielding the effective diameter, $D_{eff}$.  Caution needs to be exercised when interpreting the meaning of an effective diameter in cases where objects have high rotational amplitudes. However, all objects in our sample have peak-to-peak amplitudes less than $\sim 0.5$ magnitudes, and less than $10\%$ have peak-to-peak amplitudes greater than $\sim0.3$ magnitude. We therefore feel confident that assuming spherical shapes for our models does not significantly affect the results derived in this paper. 

Thermal fluxes were computed for each individual WISE measurement using the thermal model, ensuring that the correct Sun-observer-object geometry was used. The temperature of each facet was computed the thermal distribution assumed by the NEATM model, and color corrections were applied to each facet based on \citet{Mainzer.2011b}. In addition, adjustments of the W3 effective wavelength blue-ward by $4\%$ from $11.5608\mu$m to $11.0984\mu$m, and of the W4 effective wavelength red-ward by $2.5\%$ form $22.0883\mu$m to $22.6405\mu$m were used. Due to the red-blue calibrator discrepancy reported in \citet{Wright.2010a} and \citet{Mainzer.2011b} offsets to the W3 and W4 magnitude zeropoints of $-8\%$ and $+4\%$, respectively, were applied. In general, orbital elements and absolute magnitudes were taken from the MPC catalogs, and we assumed an error of 0.3 magnitudes for the absolute magnitude, $H$. Emissivity, $\epsilon$, was assumed to be $0.9$ for all wavelengths \citep[c.f. ][]{Harris.2009a}, and the slope parameter, G, in the magnitude-phase relationship was set to 0.15 unless an improved value exist in the MPC catalogs. 

\begin{figure}[t]
\begin{center}
\includegraphics[width=12cm]{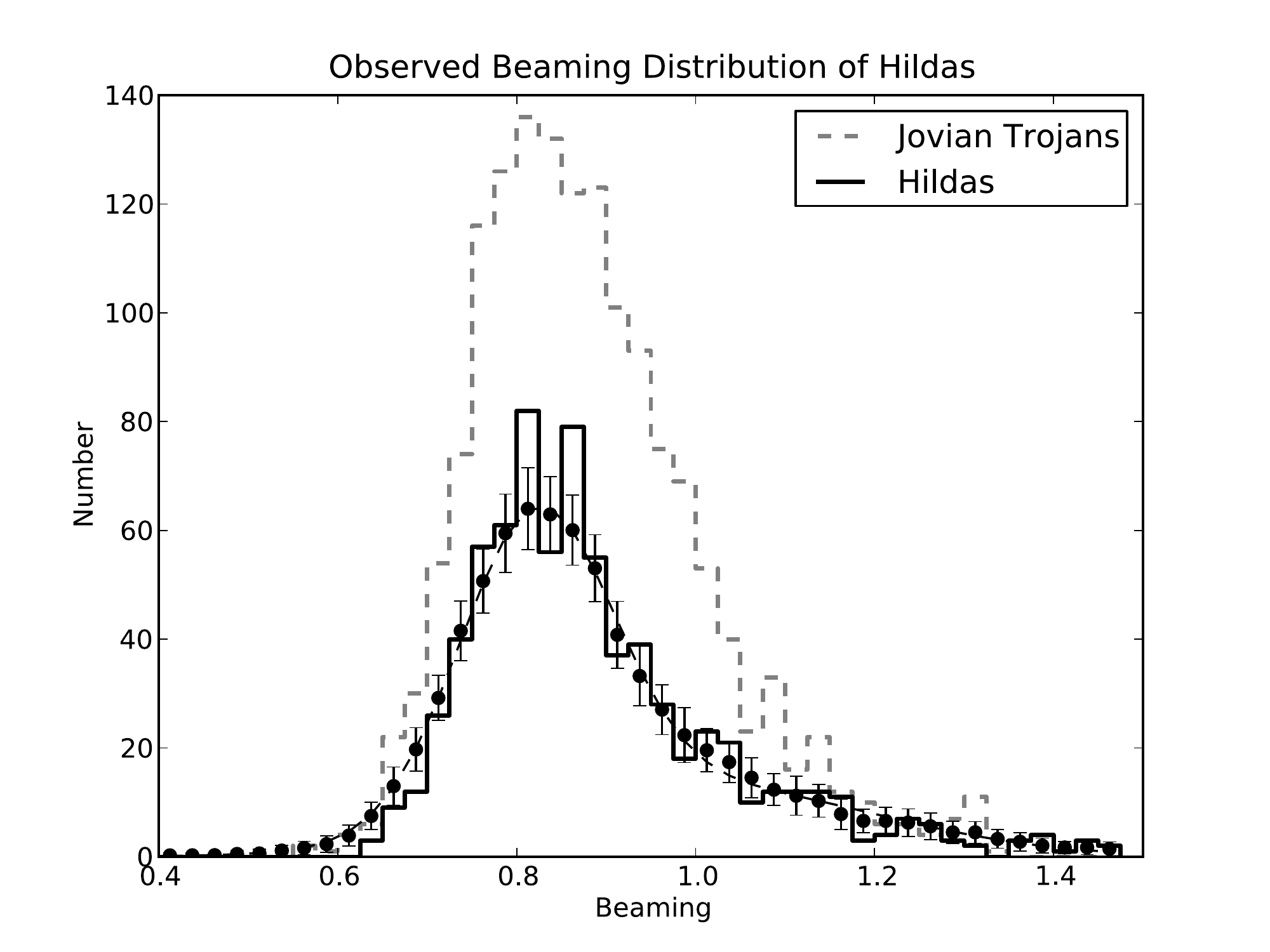}
\caption{The distribution of the beaming value, $\eta$, based on the 747 objects for which there were observations in both the W3 and W4 band. }
\label{fig:beaming}
\end{center}
\end{figure}

For Hildas with measurements in both W3 and W4, the beaming parameter $\eta$ was determined using a least square minimization, but was constrained to be less than the upper bound set by the FRM case, $\pi$. The resulting distribution based on 747 Hildas with long observational arcs is shown in Figure \ref{fig:beaming} and has a weighted average of $\eta = 0.85\pm0.12$. To understand how the errors on the derived beaming influence this distribution we employed a Monte Carlo (MC) approach. We varied the beaming for each object randomly using a Gaussian error distribution with full-width-half-max of the associated derived error. After each of the 793 objects' beaming values were varied in this fashion the distribution for each new set was computed. Using 100 such varied sets we computed the mean and associated standard deviation for each bin. The resulting distribution is shown in Figure \ref{fig:beaming} as the points with associated errorbars. The double Gaussian curve that best fits this distribution has a mean of $0.82\pm0.08$ and $1.02\pm0.19$, with the lower mean Gaussian having a peak $\sim 5$ times higher then the higher mean Gaussian.  For the objects in the LAH with only one thermal measurement, the beaming values cannot be fitted and instead are given a value $0.85\pm0.12$. The beaming distribution is similar to that of the Jovian Trojans \citep{Grav.2011b} and is slightly lower than the value of $\eta = 0.91$ derived based on 23 objects detected in two or more bands in the IRAS survey \citep{Ryan.2010a,Ryan.2011a}. 

\begin{figure}[t]
\begin{center}
\includegraphics[width=12cm]{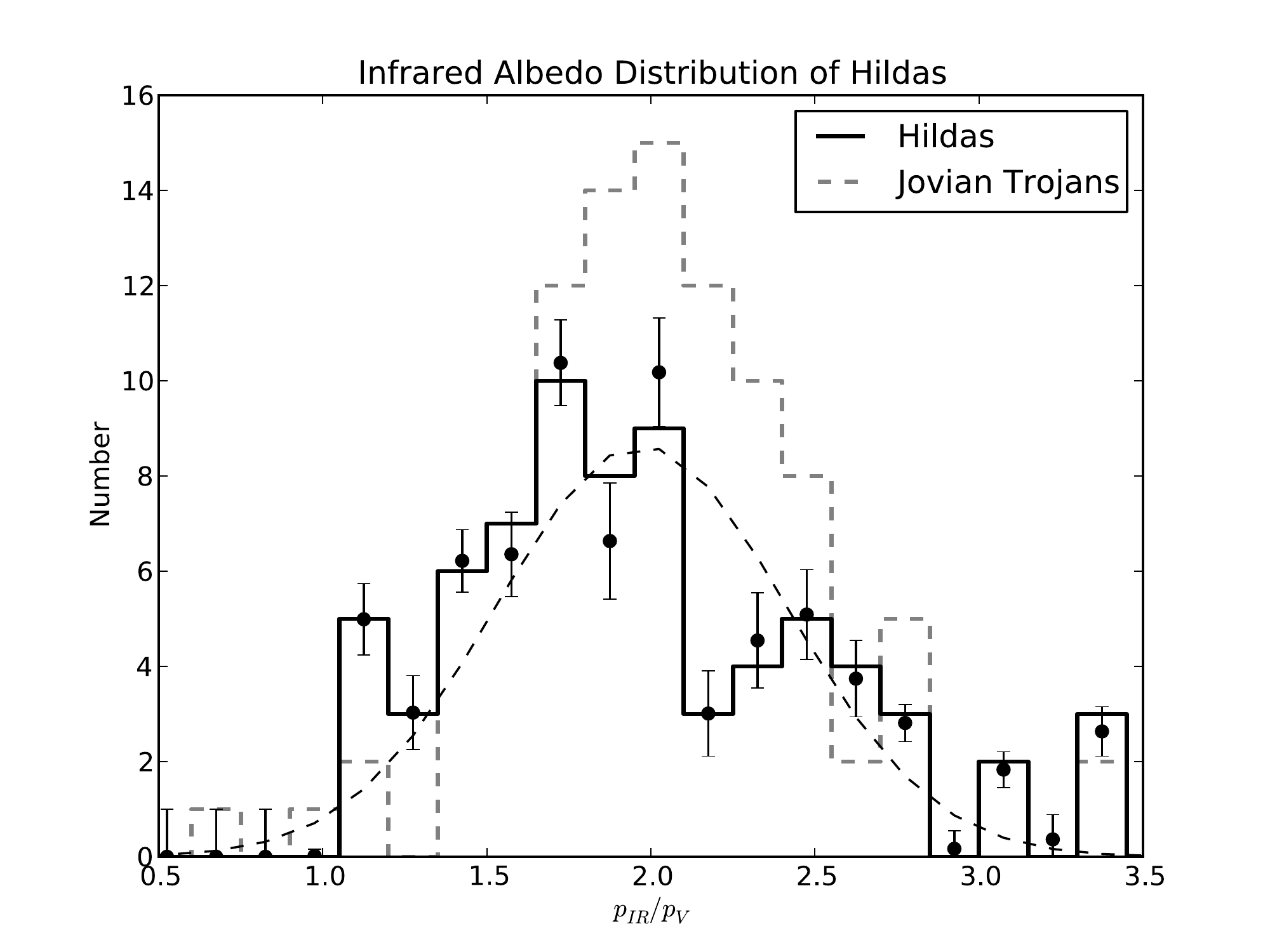}
\caption{The distribution of the $p_{IR}/p_V$ values, based on the 72 objects for which there were observations in either the W1, W2 or both bands. The points with associated errorbars is the Monte Carlo (MC) error analysis described in the text based on 100 trial runs. The best fit single Gaussian distribution, which has a mean and standard deviation of $1.9\pm0.4$, is shown as a dashed line.}
\label{fig:irfactor}
\end{center}
\end{figure}

For the Hildas bands W1 and W2 are generally dominated by reflected light. The flux due to reflected sunlight was computed for each WISE band as described in \citet{Mainzer.2011b} using the International Astronomical Union phase curve correction \citep{Bowell.1989a}. The facets that were illuminated by reflected sunlight and observable by WISE were corrected using color corrections appropriate for a G2 V star \citep{Wright.2010a}. In order to compute the fraction of total luminosity due to reflected light, the relative reflectance at bands W1 and W2, dubbed $p_{IR}/p_V$, was introduced. The distribution for $p_{IR}/p_V$ for the 72 Hildas with long observational arc that had detections in either W1, W2 or both is shown in Figure \ref{fig:irfactor}. The weighted average for the distribution is $1.9\pm0.5$, which is slightly lower than that found for the Jovian Trojans \citep{Grav.2011b}. For objects where there are no W1 and W2 observations, $p_V/p_{IR}$ is assumed to be $1.9\pm0.5$. 

\section{Results}
\label{sec:res}

\begin{figure}[t]
\begin{center}
\includegraphics[width=12cm]{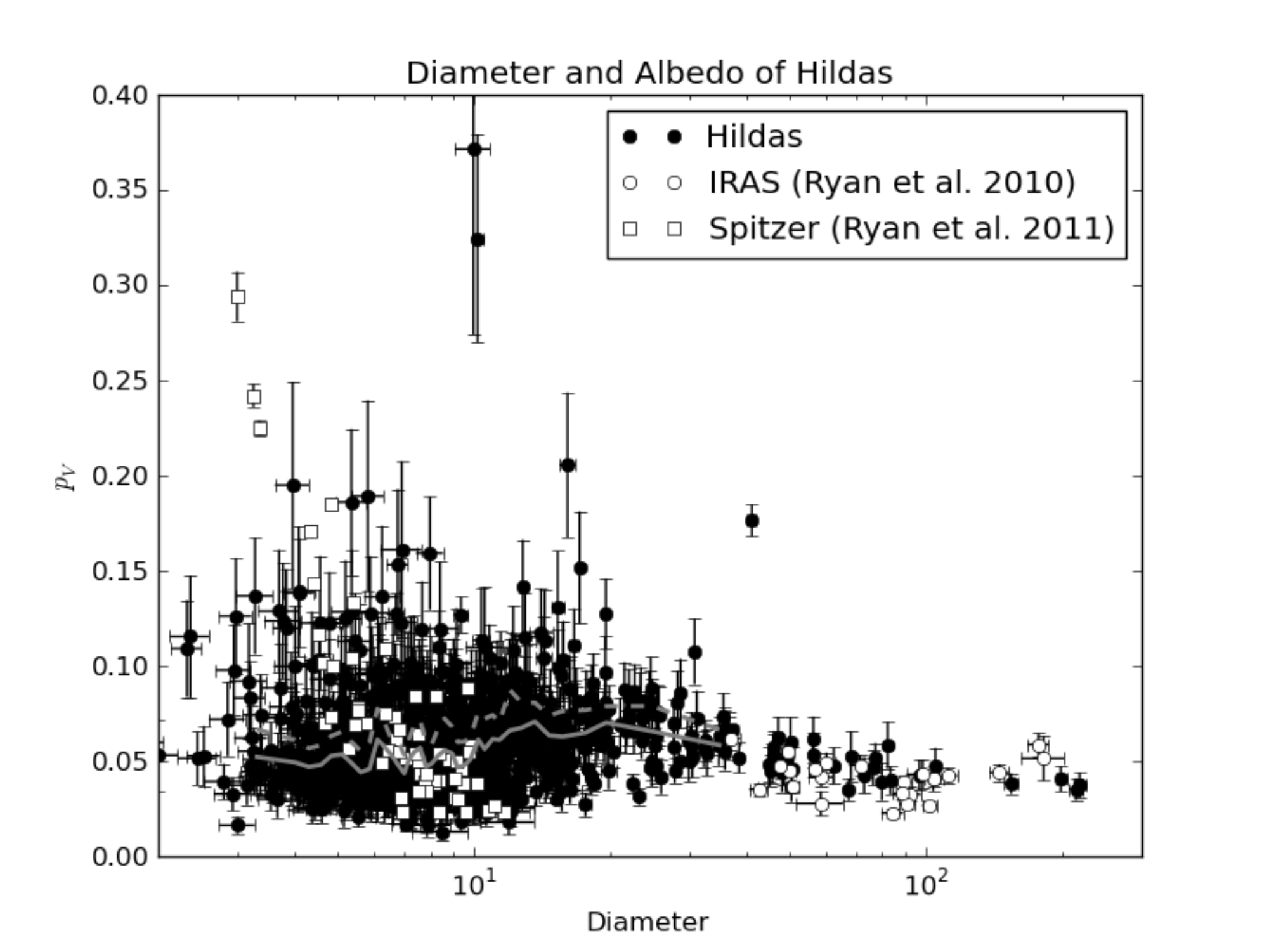}
\caption{The diameters and albedos of the 1023 Hildas with long observational arc for which a thermal model was derived are shown. The values for the 23 Hildas observed by IRAS \citep{Ryan.2010a} and 64 objects observed by Spitzer \citep{Ryan.2011a} are also plotted. The calculation of the running median and the absolute median deviation is shown in grey.}
\label{fig:DpV}
\end{center}
\end{figure}

\begin{figure}[t]
\begin{center}
\includegraphics[width=12cm]{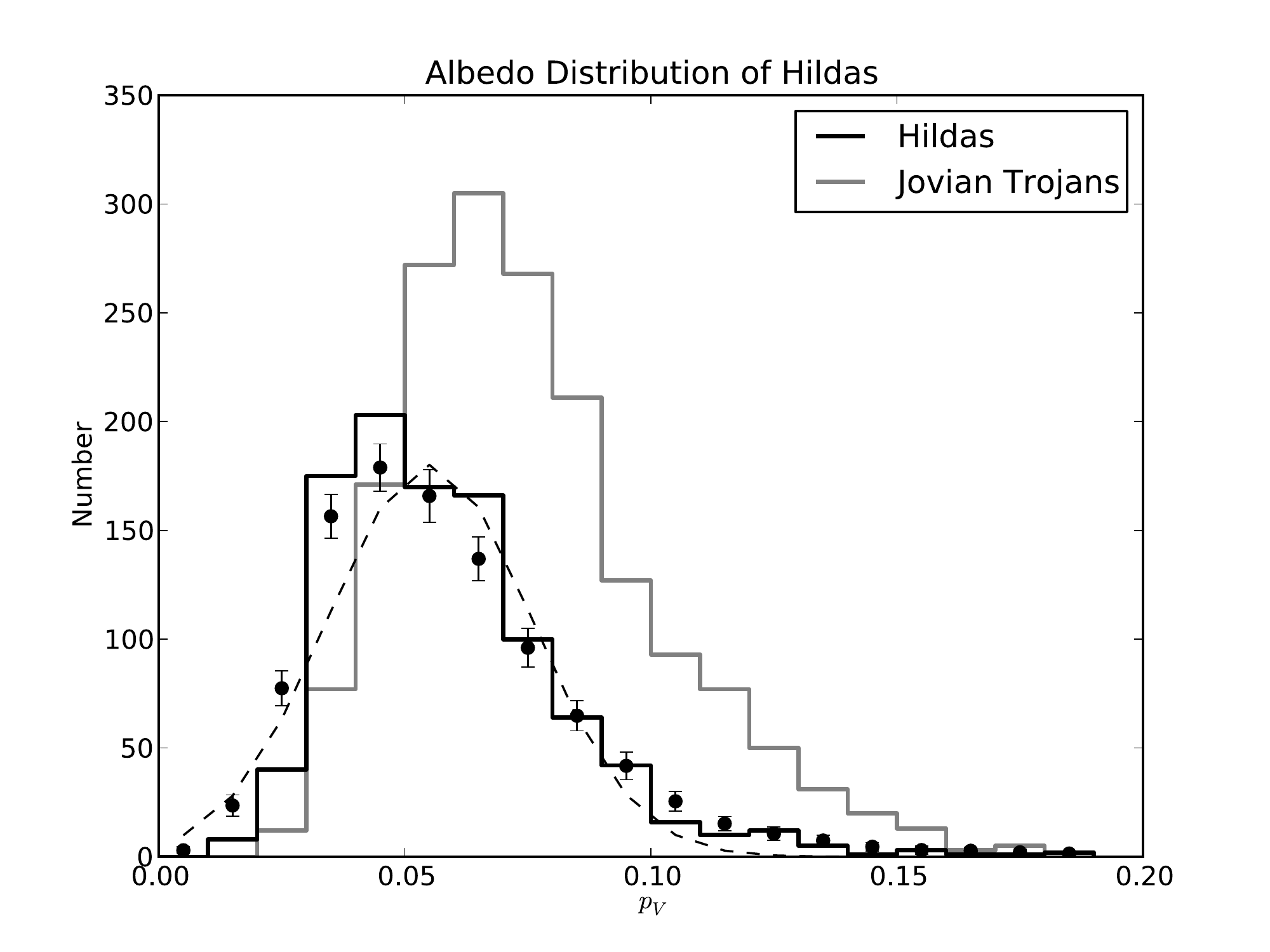}
\caption{The albedo distribution of the 1023 Hildas for which a thermal model was derived are shown. The distribution using the MC error analysis as described in Section \ref{sec:thermal} is shown as black points with associated error bars. A best fit single Gaussian distribution with mean and standard deviation of $0.055\pm0.021$ is shown as a dashed line.}
\label{fig:albedo}
\end{center}
\end{figure}

We were able derive diameters and albedos for 1023 of the Hildas in the LAH sample. The results are plotted in Figure \ref{fig:DpV}, together with the 23 objects observed by IRAS \citep{Ryan.2010a} and 64 objects observed by Spitzer \citep{Ryan.2011a}. The albedo distribution is homogeneous and very low with a weighted mean of $0.055\pm0.018$ (see Figure \ref{fig:albedo}).  There is only a handful of objects with higher albedo, $p_V >  0.15$, that would be indicative of high-albedo interlopers into a generally dark population. The albedo distribution is clearly darker than the Jovian Trojan population \citep{Grav.2011b}. We caution that although there does appear to be a broadening of the albedo distribution for smaller sizes, this does not mean that there is a correlation between size and albedo as reported by other authors \citep{Ryan.2011a}. We computed the running median and running median absolute deviation across the size range observed using a variety of window sizes and found that both the median and deviation remain consistent with the weighted mean across the full size range. The maximum median of $0.065\pm0.008$ is found around $20$km and then decreases slightly to $0.047\pm0.010$ at 4km. The broadening seen at smaller sizes it thus a natural increase in the number of outlier measurements following Gaussian errors due to the increase of the total number of objects at smaller sizes, not a broadening due to some physical difference or process happening on the surface.

\begin{figure}[t]
\begin{center}
\includegraphics[width=12cm]{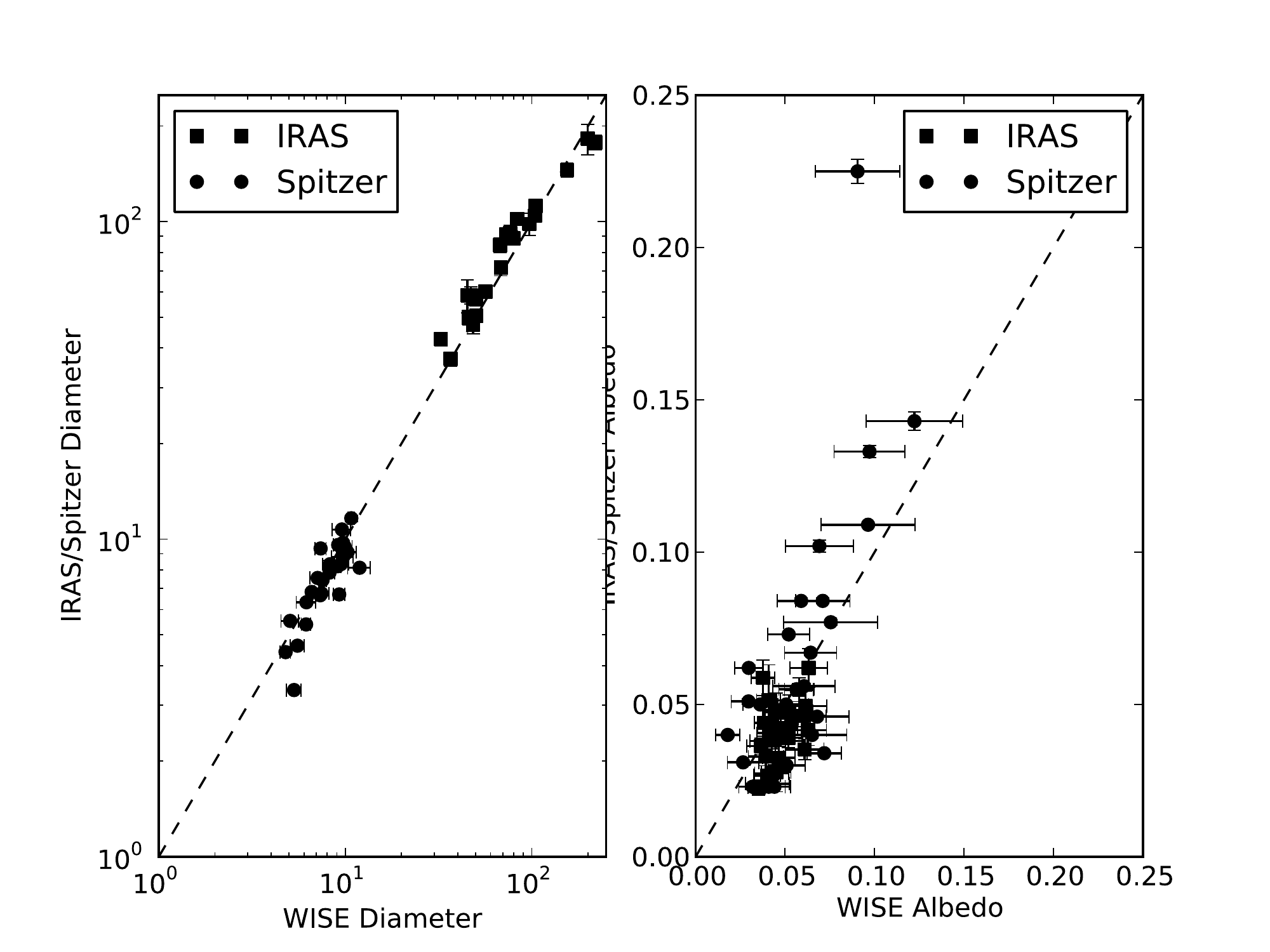}
\caption{The comparison of the derived diameters (left panel) and albedo (right panel) from this paper and those from \citet{Ryan.2010a,Ryan.2011a}.}
\label{fig:compare}
\end{center}
\end{figure}

Figure \ref{fig:compare} shows the albedo and diameter from the IRAS and Spitzer samples compared to that from our sample. It is seen that the IRAS diameters derived by \citet{Ryan.2010a} are systematically slightly larger than our values, a result that was also seen in comparison of the diameters of the other populations \citep{Mainzer.2011d}. The diameters and albedos derived by \citet{Ryan.2011a} from their Spitzer survey are generally in very good agreement with the values derived here, although there are small systematic shifts with the objects being slightly smaller and darker in their results. NEOWISE observed one of the five high albedo objects, (128295) 2003 WD111, used by \citet{Ryan.2011a}  to argue for the size-albedo dependency. This object only has an albedo of $p_V = 0.09\pm0.02$ in our data, less than half the value reported in that paper. It should be noted here that the uncertainties quoted in Table 2 of \citet{Ryan.2011a} (plotted in Figure \ref{fig:compare}) are understated as Spitzer $24\mu$m MIPS photometric observation have a minimum calibration uncertainty of $4\%$ according to the Spitzer Instrument Handbook. While back of the envelope error calculations from uncertainties in H (using $\pm0.1$ magnitude uncertainty, rather than the more realistic $\pm0.3$ magnitude uncertainty used in this paper) and the beaming $\eta$ are presented in their paper, these uncertainties were not folded into the table of derived values they presented. This makes it difficult to accurately determine why and if the derived values in this paper and that of \citet{Ryan.2011a} are significantly different. The WISE results have been extensively calibrated against asteroids with known diameter in \citet{Mainzer.2011b} and a comparison with the IRAS sample is found in \citep{Mainzer.2011d}. The latter paper shows that the IRAS-based diameter values derived by \citet{Ryan.2010a} appears to be systematic larger than diameters from radar, occultation or spacecraft flybys. This systematic offset is not seen nearly as strongly in the original IRAS catalog by \citet{Tedesco.2002a}. The validity of the claim of an albedo dependency with diameter put forth in \citet{Ryan.2011a} will be studied closer in Section \ref{sec:sizefreq} below.

\begin{figure}[t]
\begin{center}
\includegraphics[width=12cm]{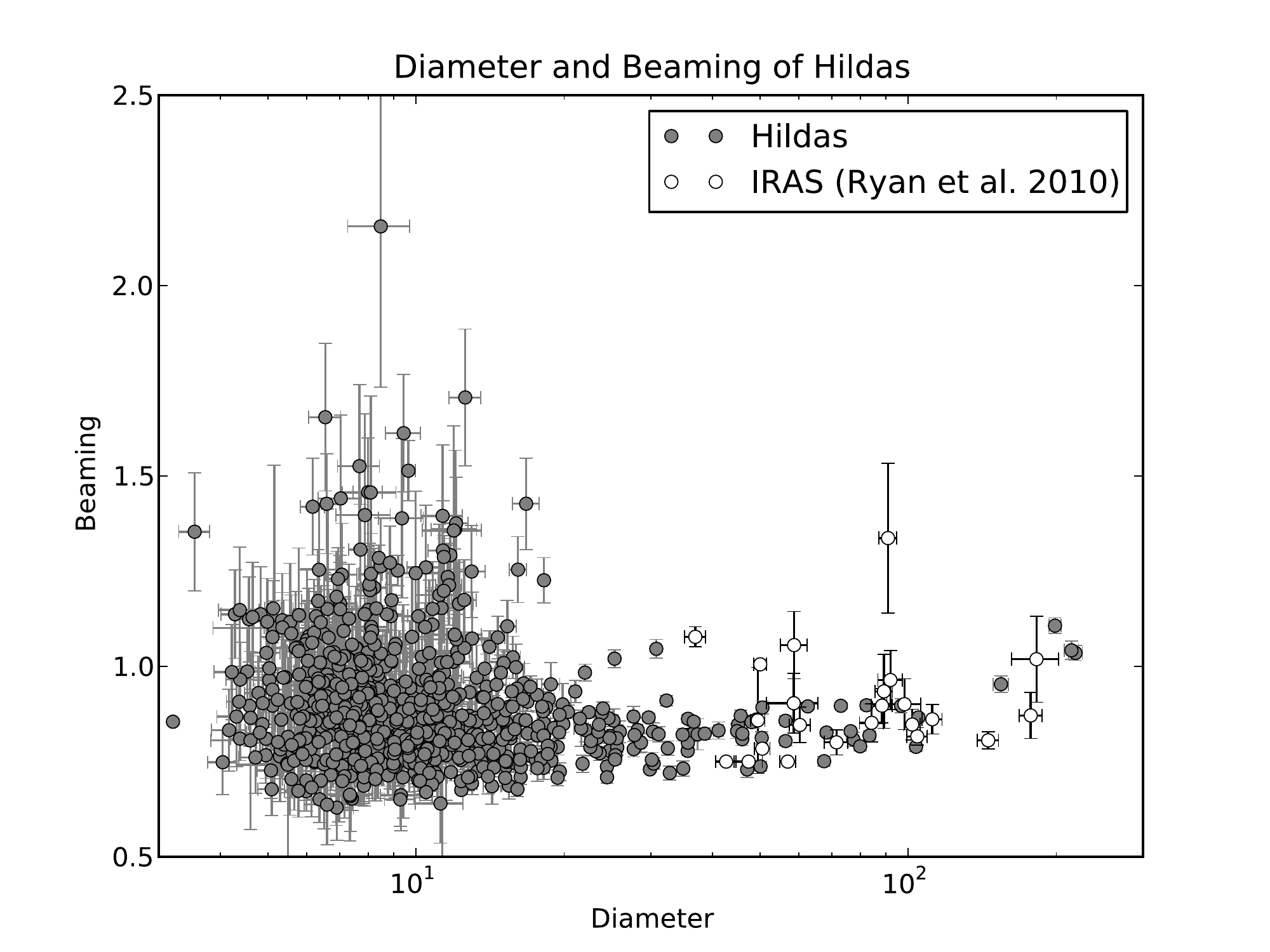}
\caption{Shown are the diameters and beaming values of the 747 Hildas with long observational arcs for which a thermal model with varied beaming value was derived. The values for the 23 Hildas observed by IRAS \citep{Ryan.2010a} are also shown for comparison.}
\label{fig:Deta}
\end{center}
\end{figure}

The diameter and beaming values for the 747 Hildas with long observational arcs that had observations in two thermal bands are shown in Figure \ref{fig:Deta}. The beaming is generally centered around the weighted mean of $0.85\pm0.12$, although there is a small increase of higher beaming values at smaller sizes. This is most likely a result of the increasing number of objects at smaller sizes, resulting in more outlier objects in the wings of the beaming distribution, rather than a real physical widening of beaming values for the population.

\subsection{Dual Epoch Objects}
\begin{figure}[t]
\begin{center}
\includegraphics[width=12cm]{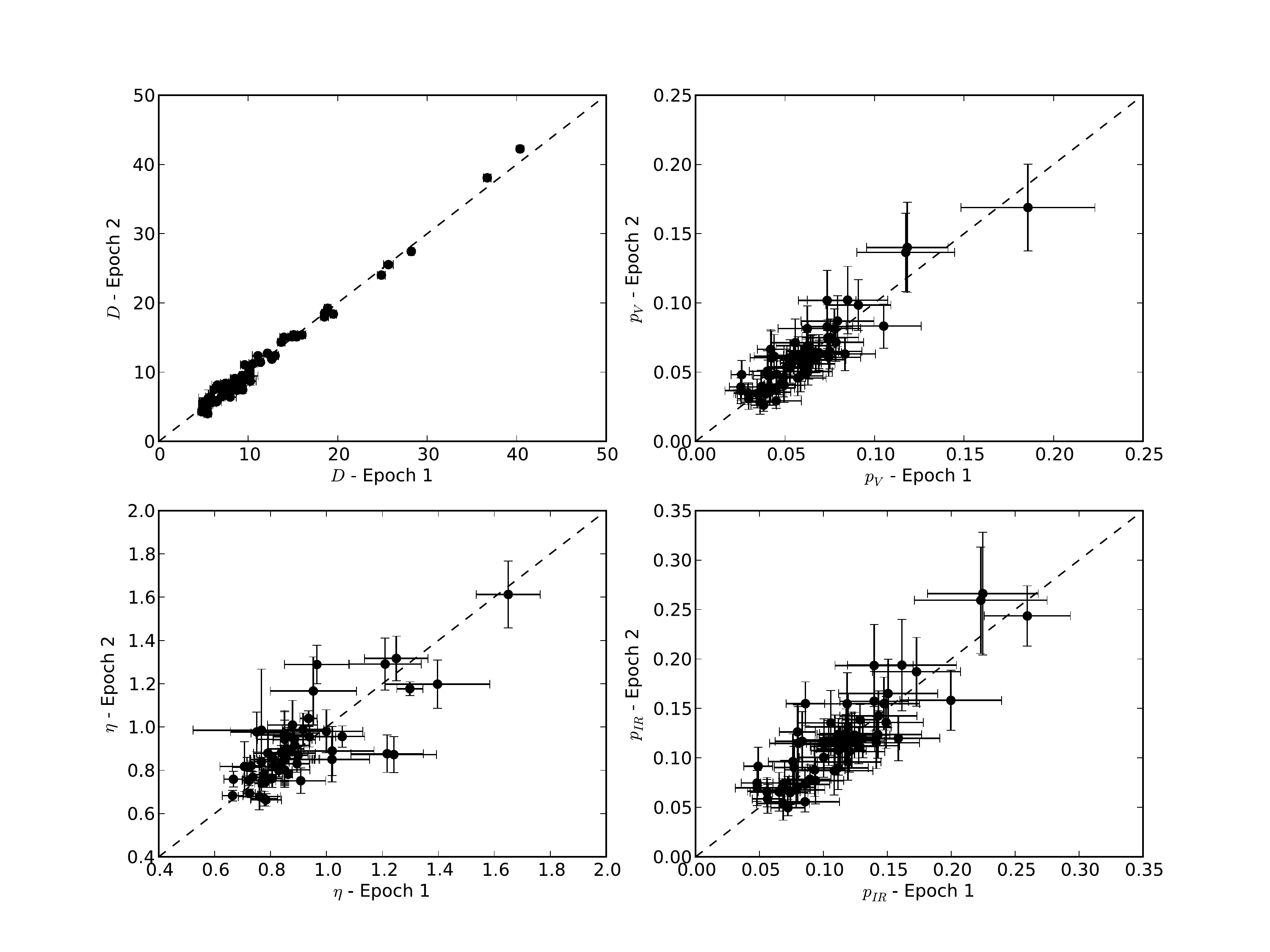}
\caption{Comparison of the thermal model variables for the 66 objects for which WISE observed at two different epoch during the cryogenic survey.}
\label{fig:epoch2}
\end{center}
\end{figure}

There are 66 objects in our sample that NEOWISE observed at two different epochs during the cryogenic survey.  For each of these objects, each epoch was fitted independently. The results are shown in Figure \ref{fig:epoch2}, and it is seen that for all derived parameters the difference between the values in the two epochs are consistent to within the derived errors. 

\subsection{High Albedo Objects}

There are 8 objects with $p_V > 0.17$ that could be higher albedo interlopers into a generally dark Hilda population (see Table \ref{tab:highalb}. Of the 8 only one, (3290) Azabu, has SDSS photometry \citep{Gil-Hutton.2008a} and none have any spectral observations. \citet{Gil-Hutton.2008a} identified (3290) Azabu as an X-complex asteroid, and the high albedo found here $p_V = 0.32\pm0.08$ would make it an E-type asteroid.

\begin{table}[htdp]
\caption{High Albedo Hildas}
\begin{center}
\begin{tabular}{cccc}
Object & Diameter & Beaming & Albedo \\
& [km] & & \\
\hline
1162 & $41.3\pm0.9$ & $0.83\pm0.03$ & $0.18\pm0.03$  \\
3290 & $10.2\pm0.4$ & $0.70\pm0.06$ & $0.32\pm0.08$  \\
11249 & $10.0\pm0.9$ & $0.86\pm0.14$ & $0.37\pm0.10$  \\
14699 & $16.1\pm0.7$ & $1.25\pm0.09$ & $0.21\pm0.04$  \\
77734 & $5.4\pm0.4$ & & $0.19\pm0.04$  \\
89928 & $5.8\pm0.5$ & $0.77\pm0.11$ & $0.19\pm0.05$  \\
96086 & $6.3\pm0.5$ & & $0.19\pm0.05$  \\
225800 & $4.0\pm0.3$ & & $0.19\pm0.05$  \\
\end{tabular}
\end{center}
\label{tab:highalb}
\end{table}%

\subsection{Taxonomy}

\begin{figure}[t]
\begin{center}
\includegraphics[width=12cm]{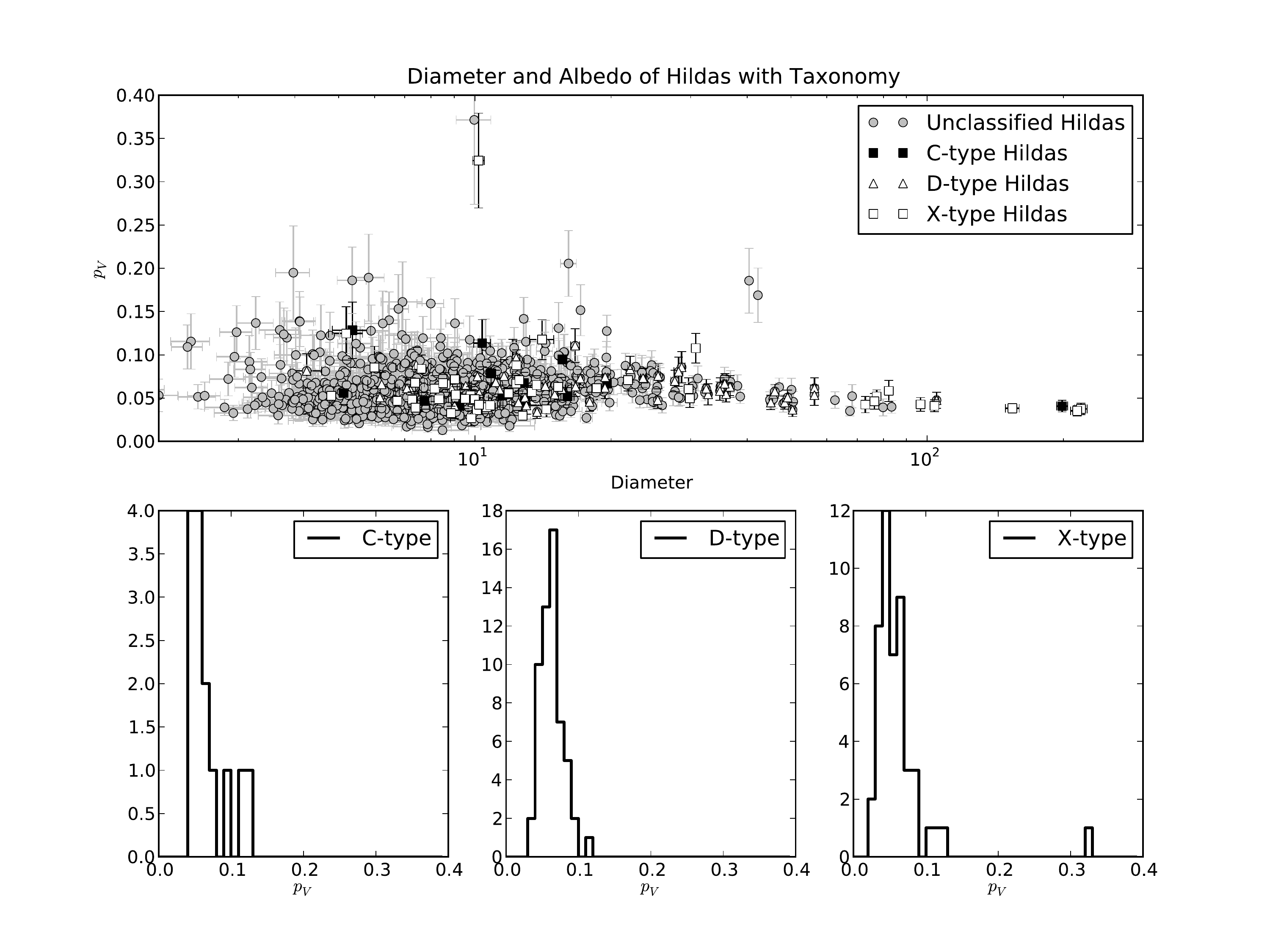}
\caption{Shown here is the diameter and $p_{V}$ of the Hildas with both unclassified and classified taxonomy from other authors \citep{Dahlgren.1995a,Dahlgren.1997a,Bus.2002a,Xu.1995a,Lazzaro.2004a,Gil-Hutton.2008a}.}
\label{fig:DpVtax}
\end{center}
\end{figure}

Several authors have classified a number of the Hildas in the Tholen taxonomy scheme, based either on multi-color photometric \citep{Gil-Hutton.2008a} or visible wavelength spectroscopic observations \citep{Bus.2002a,Dahlgren.1995a,Dahlgren.1997a,Lazzaro.2004a,Xu.1995a,Fornasier.2011a}. We found 123 objects among our sample that have taxonomic classes assigned by these authors, and these are plotted in Figure \ref{fig:DpVtax}. 

\begin{figure}[t]
\begin{center}
\includegraphics[width=12cm]{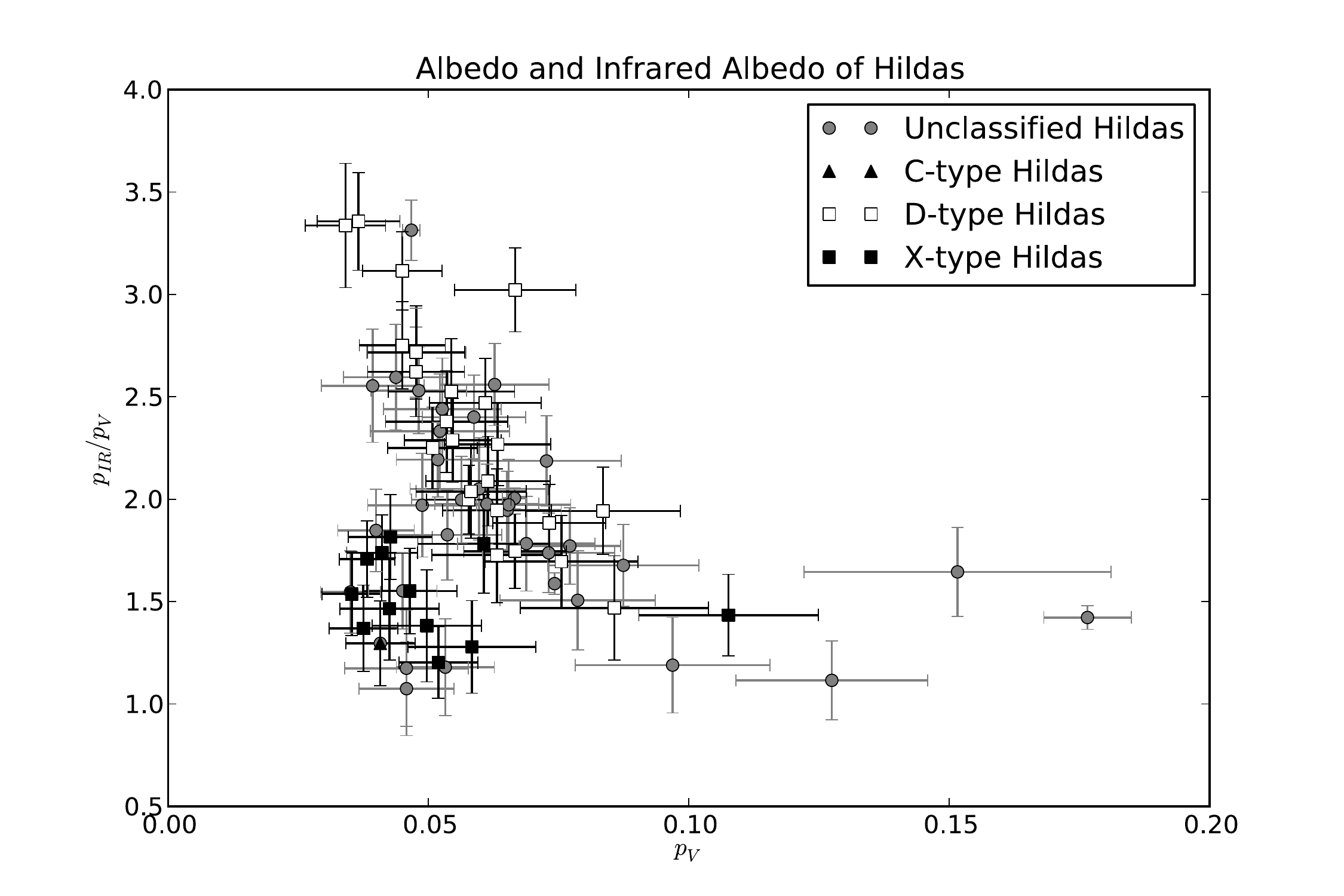}
\caption{The visible albedo $p_V$ and relative reflectance in the W1/W2 bands, $p_{IR}/p_V$ are shown. Also shown is the taxonomic classifications based on data from literature \citep{Dahlgren.1995a,Dahlgren.1997a,Bus.2002a,Xu.1995a,Lazzaro.2004a,Gil-Hutton.2008a,Fornasier.2011a}.}
\label{fig:pVpIR}
\end{center}
\end{figure}

As mentioned in Section \ref{sec:thermal}, the survey yielded 71 objects with observations in either W1, W2 or both, making it possible to derive the relative reflectance in the W1/W2 bands. Figure \ref{fig:pVpIR} shows the albedo, $p_V$, versus this relative reflectance for these 71 objects. The spectral class for the 36 objects in this sample that have been studied among the photometric or spectral surveys mentioned above are also given. It is seen that the C- and X-type objects cluster in the group having both low $p_V$ and $p_{IR}/p_V$, while the objects classified as D-type are all in the group with low $p_V$ and moderate $p_{IR}/p_V$. The low albedos of the X-complex objects indicate that these are P-types, rather than the moderate albedo M-types or high albedo E-types. The larger range of albedos seen here in the D-type group ($p_V \sim 0.03 - 0.10$) compared to the C- and P-type group ($p_V \sim 0.03-0.06$) is consistent with the albedo distribution of these types as seen in the MBAs \citep{Mainzer.2011d}. We note that there is no apparent way with our data alone to distinguish between between the C- and P-type objects in the Hilda population. 

From Figures \ref{fig:DpV} and \ref{fig:pVpIR} it is seen that the large objects are all C- or P-type asteroids, with the largest D-type object, (1269) Rollandia, being $D \sim 104\pm1$km. This confirms the result of other surveys \citep{Dahlgren.1997a}. As we will see later in Section \ref{sec:sizefreq} it is clear that the known sample is more than $90\%$ complete for sizes of 10km or larger. Looking at our sample, there is only one object with diameter larger than 30km for which the $p_{IR}/p_V$ were not derived, and there are 49 objects in our sample of this size or larger. Of these, 13 land in the C- or P-type grouping in Figure \ref{fig:pVpIR}, while 33 fall in the D-type grouping. Two of the objects are consistent with M-type classification. The faintest of the objects with diameter larger than 30km is (5928) with $H = 11.4$. There are 16 objects with $H \leq 11.4$ that are not in the LAH sample, and two of these have taxonomy determined by other sources (one X-type and one D-type). This means that for objects with diameter larger than 30km the fraction of of C-/P-types is  $26^{+17}_{-5}\%$, while the D-type fraction $67^{+7}_{-15}\%$. The fraction of C-/P-types are of course dominated by P-types, with (334) Chicago being the only well defined C-type object among the Hildas and (1439) Vogtia having a possible F-type classification.  

Two of the objects as X-type ((3843) OISCA and (11542) 1992 SU21) are located among D-types. One of these, (11542) 1992 SU21, is classified based on SDSS photometry by \citet{Gil-Hutton.2008a}, while the other, (3843) OISCA, was generically given E, M, or P as the possible Tholen classification by \citet{Dahlgren.1997a}. The albedo of (3843) OSICA is $p_V = 0.11\pm0.01$, which would make this object an M-type asteroid. This classification leads us to believe that the objects in Figure \ref{fig:pVpIR} with $p_V > 0.1$ may all be M-type objects, due to their low slopes (i.e. low $p_{IR}/p_v$) and moderate visible albedos. Additional spectral observations are needed to confirm these classifications or possibly reclassify them as D-type asteroid as indicated by their location in Figure \ref{fig:pVpIR}. 

\begin{table}[htdp]
\tabletypesize{\scriptsize}
\caption{New or reclassified taxonomy of the objects in the Hilda population based on the visible albedo and relative reflectance in bands W1/W2.}
\begin{center}
{\scriptsize
\begin{tabular}{ccccc}
 & \multicolumn{2}{c}{Taxonomy} & \\
Object & Old & New & Albedo & $p_{IR}/p_V$ \\
\hline
1162 & & M & $0.18\pm0.04$ & $1.42\pm0.24$ \\
1256 & & D & $0.05\pm0.01$ & $2.33\pm0.22$ \\
1439 & & C or P & $0.05\pm0.01$ & $1.17\pm0.26$ \\
1578 & & D & $0.06\pm0.01$ & $2.56\pm0.20$ \\
1746 & & D & $0.05\pm0.01$ & $2.72\pm0.22$ \\
1748 & & D & $0.05\pm0.01$ & $2.53\pm0.21$ \\
1877 & & D & $0.07\pm0.01$ & $1.74\pm0.19$ \\
1911 & & C or P & $0.04\pm0.01$ & $1.55\pm0.20$ \\
1941 & & M & $0.15\pm0.03$ & $1.65\pm0.22$ \\
2067 & & D & $0.05\pm0.01$ & $1.83\pm0.22$ \\
2312 & & D & $0.06\pm0.01$ & $2.05\pm0.25$ \\
3254 & & D & $0.07\pm0.01$ & $2.19\pm0.22$ \\
3290 & X & E & $0.32\pm0.08$ & $1.90\pm0.34$ \\
3843 & X & M & $0.11\pm0.01$ & $1.43\pm0.21$ \\
4196 & & D & $0.07\pm0.01$ & $2.02\pm0.23$ \\
4317 & & D & $0.05\pm0.01$ & $2.19\pm0.18$ \\
5603 & & D & $0.05\pm0.01$ & $1.97\pm0.25$ \\
5928 & & D & $0.05\pm0.01$ & $2.44\pm0.25$ \\
6984 & & D & $0.04\pm0.01$ & $2.60\pm0.26$ \\
7027 & & D & $0.07\pm0.01$ & $1.95\pm0.19$ \\
7174 & & D & $0.07\pm0.01$ & $1.59\pm0.22$ \\
8550 & & C or P & $0.05\pm0.01$ & $1.07\pm0.23$ \\
8915 & & D & $0.06\pm0.01$ & $2.40\pm0.20$ \\
10331 & & M & $0.13\pm0.02$ & $1.12\pm0.19$ \\
11542 & X & D & $0.06\pm0.01$ & $1.78\pm0.17$ \\
13035 & & C or P & $0.05\pm0.01$ & $1.30\pm0.22$ \\
15231 & & D & $0.06\pm0.01$ & $1.98\pm0.19$ \\
15376 & & D & $0.08\pm0.01$ & $1.51\pm0.24$ \\
15638 & & D & $0.06\pm0.01$ & $2.00\pm0.21$ \\
20038 & & D & $0.08\pm0.01$ & $1.77\pm0.19$ \\
31817 & & D & $0.09\pm0.01$ & $1.68\pm0.20$ \\
32460 & & M & $0.10\pm0.02$ & $1.19\pm0.23$ \\
38613 & & D & $0.05\pm0.01$ & $3.19\pm0.24$ \\
47907 & & D & $0.07\pm0.01$ & $1.78\pm0.23$ \\
61042 & & D & $0.07\pm0.01$ & $1.97\pm0.22$ \\
\end{tabular}}
\end{center}
\label{tab:tax}
\end{table}%

\clearpage

\subsection{Hilda and Schubart Collisional Families}

\citet{Broz.2008a} have identified two potential collisional families in the Hilda populations, one centered on (153) Hilda and the other on (1193) Schubart. 

\begin{figure}[t]
\begin{center}
\includegraphics[width=12cm]{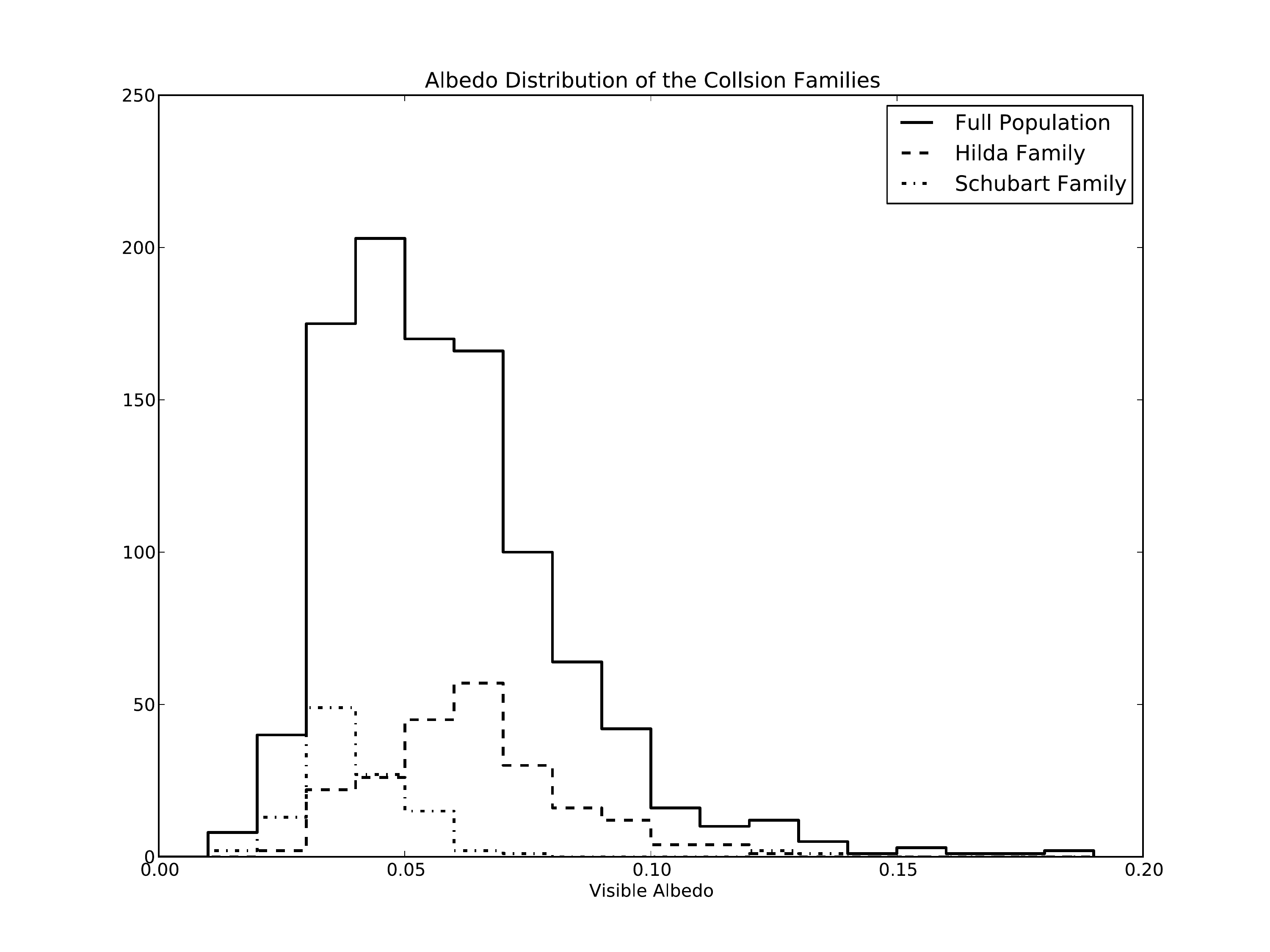}
\caption{The visible albedo $p_V$ of the two known collisional families identified by \citet{Broz.2008a} in the Hilda population.}
\label{fig:pVfam}
\end{center}
\end{figure}

We observed 219 of the 360 members identified by \citet{Broz.2008a}\footnote{Lists of members of the Hilda and Schubart collisional families were taken from the home page of M. Bro{\v z} at http://sirrah.troja.mff.cuni.cz/$\sim$mira/mp/trojans\_hildas/} and their weighted mean albedo is $0.061\pm0.011$. The albedo distribution is shown in Figure \ref{fig:pVfam} and is seen to be brighter than the full Hilda population.  Our value is almost $50\%$ brighter than the value used by \citet{Broz.2008a} and \citet{Broz.2011a} in attempts to derive the size of the parent body of this collisional family. 
13 of the objects in this family have $p_{IR}/p_V$ derived, and all except for two, (153) Hilda and (65374) 2002 PP55, are found to be in the D-type cluster in Figure \ref{fig:pVpIR}. This is contrary to that stated in \citet{Broz.2011a} which claim that most objects in the Hilda collisional family are C-type objects based on the spectral slopes derived from SDSS photometry \citep{Ivezic.2002a,Parker.2008a,Gil-Hutton.2008a}. 

For the Schubart collisional family we observed 112 out of the 232 objects identified by \citet{Broz.2008a} as members. The weighted mean of this family is $0.039\pm0.013$ and is clearly darker than that of the Hilda family (see Figure \ref{fig:pVfam}). Four of the objects observed had flux in either W1, W2 or both allowing the relative reflectance, $p_{IR}/p_V$. Following the discussion above on classification of objects, all four are found to be in the C- and P-type cluster in Figure \ref{fig:pVpIR}. 

\subsection{The Size-Frequency Distribution}
\label{sec:sizefreq}

One of the important questions regarding the Hilda populations is its size and albedo distribution. \citet{Ryan.2011a} found a significant size-albedo correlation, where smaller objects have significant higher albedo than the larger objects. It is important to note that when they applied this size-albedo relation to the known sample to derive the size-frequency relationship, their results showed a very shallow slope for the objects in 5 to 12 kilometer range. We believe that \citet{Ryan.2011a} incorrectly used the assumption that the optical surveys are currently complete to $V\sim21.5$. Currently, optical surveys like Catalina Sky Survey and Pan-STARRS \citep{Wainscoat.2010a} routinely report new discoveries in the MBA, Hildas and Jovian Trojans that are brighter than $V\sim 21.5$. \citet{Ryan.2011a} translated this assumption to a completeness for the Hilda population of $H \sim 15$. A quick  look at the MPC orbital database reveals that there are $\sim 1389$ known objects in the Hilda population with $H < 15$ and $50$ of these, making up $\sim 4-5\%$ of the known sample with $H < 15$, were discovered in the last two years. This shows that there most certainly is a low, but non-negligible, fraction of objects with $H < 15$ among the Hilda population that have yet to be discovered. 

This problem is minor, compared to the assumption that there is an albedo-size dependency, with the albedo increasing significantly for smaller sizes, which does not seem to be supported by our results. It could, however, be that our survey is simple significantly less sensitive to the higher albedo objects. In order to test this we have developed a survey simulator that mimics the real survey performed by NEOWISE and it is briefly described in \citet{Grav.2011b} and discussed in detail in \citet{Mainzer.2011f}. A synthetic population of the Hildas was generated based on \citet{Grav.2011a}, assuring that the main feature of the orbital distribution was retained. The most complex feature to duplicate is the triangular shape formed by the Hilda population (with its corners at $\pm 60$ and 180 degrees away from Jupiter in its orbit), but this was easily accomplished by remembering that the Hildas follow the librating critical argument $\sigma = 3 \lambda_J - 2 \lambda - \bar{\omega}$, where $\lambda_J$ is the mean longitude of Jupiter, $\lambda$ is the mean longitude of the asteroid and $\bar{\omega}$ is the longitude of perihelion of the asteroid \citep{Broz.2011a}. For example, if the mean longitude with respect to Jupiter is $\pm 60^\circ$ or $180^\circ$, i.e. at the Jovian Trojan clouds or opposite the Sun from Jupiter, the object has to be at aphelion in its orbit, $M \sim 180^\circ$. Objects that are half way in mean longitude between these three corners are at the perihelion point of their orbits, $M \sim 0^\circ$. 

\begin{figure}[t]
\begin{center}
\includegraphics[width=12cm]{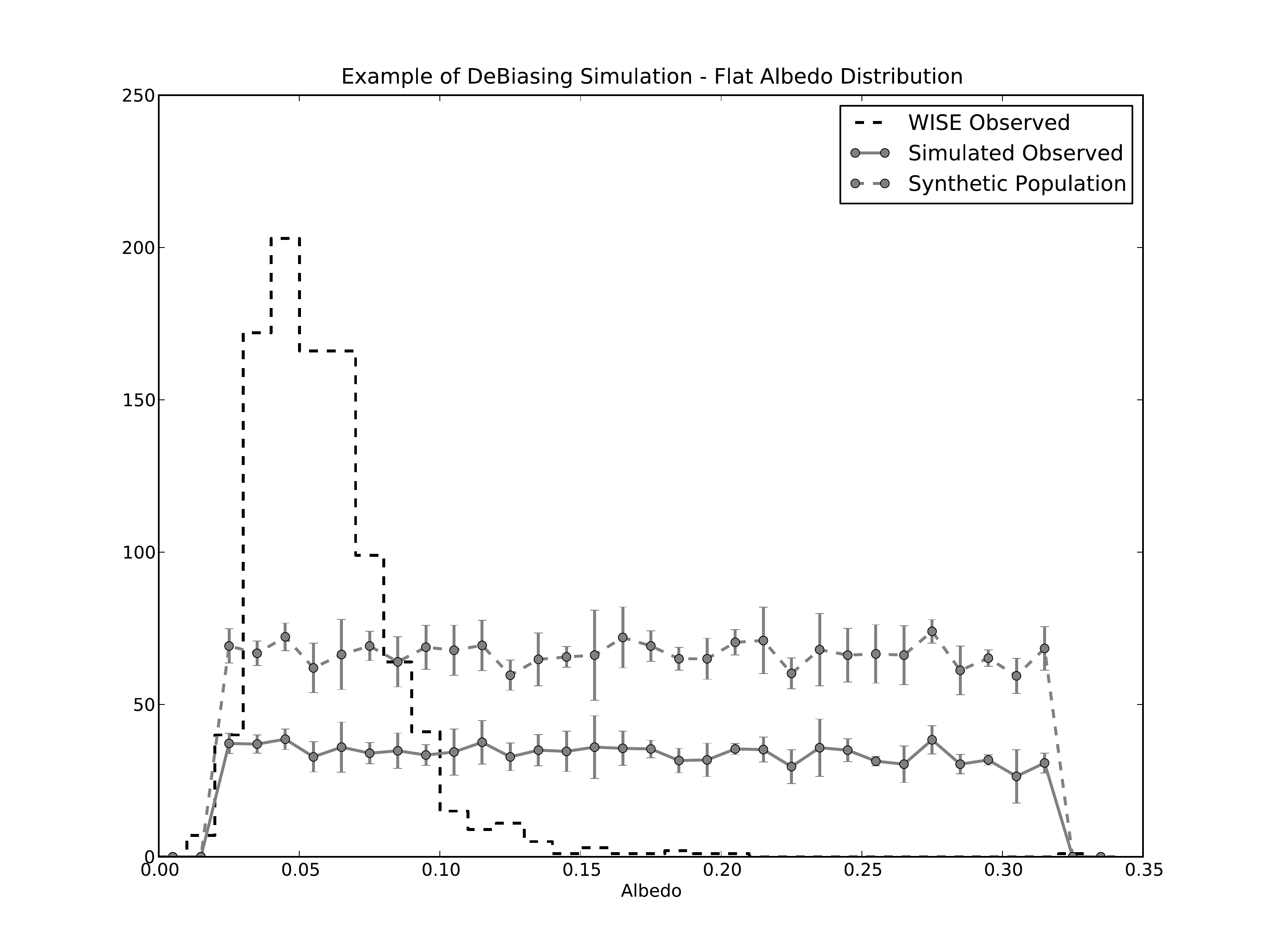}
\caption{The visible albedo, $p_V$, distributions of the synthetic population and simulated observed sample using a flat albedo distribution are shown. The albedo bias is given by dividing the simulated distribution by the synthetic, which for the albedo yields essentially a flat bias. This means that the NEOWISE survey is equally sensitive to low and high albedos. The distribution of the albedos of the Hilda population as detected by NEOWISE is shown for comparison. }
\label{fig:debias1}
\end{center}
\end{figure}

First we examine the bias that exists in the survey with respect to albedo. We use our Hilda synthetic population and assign each object a set of physical parameters. To test the albedo bias we use random distribution of albedo ranging from $2-32\%$ for each object; the beaming was given as a Gaussian distribution with mean and standard deviation of $0.85\pm0.12$. We tested both the size-frequency distribution given by \citet{Ryan.2011a} as well as a single power-law of $N(> D) \sim D^{-\alpha}$ with slope of $\alpha = 2.0$. The result was the same in all simulations, showing no significant bias for all values of albedo (see Figure \ref{fig:debias1}). This strengthens our concern that the size-albedo distribution reported by \citet{Ryan.2011a} is erroneous and a result of an observational bias caused by selecting the Spitzer targets from objects discovered solely by visible light surveys, which preferentially select against small, low albedo objects. Note that our sample from NEOWISE does not suffer from such selection biases as it is essentially a {\it blind} survey, using no apriori information in searching the data sets for both known and new minor planets. 

\begin{figure}[t]
\begin{center}
\includegraphics[width=12cm]{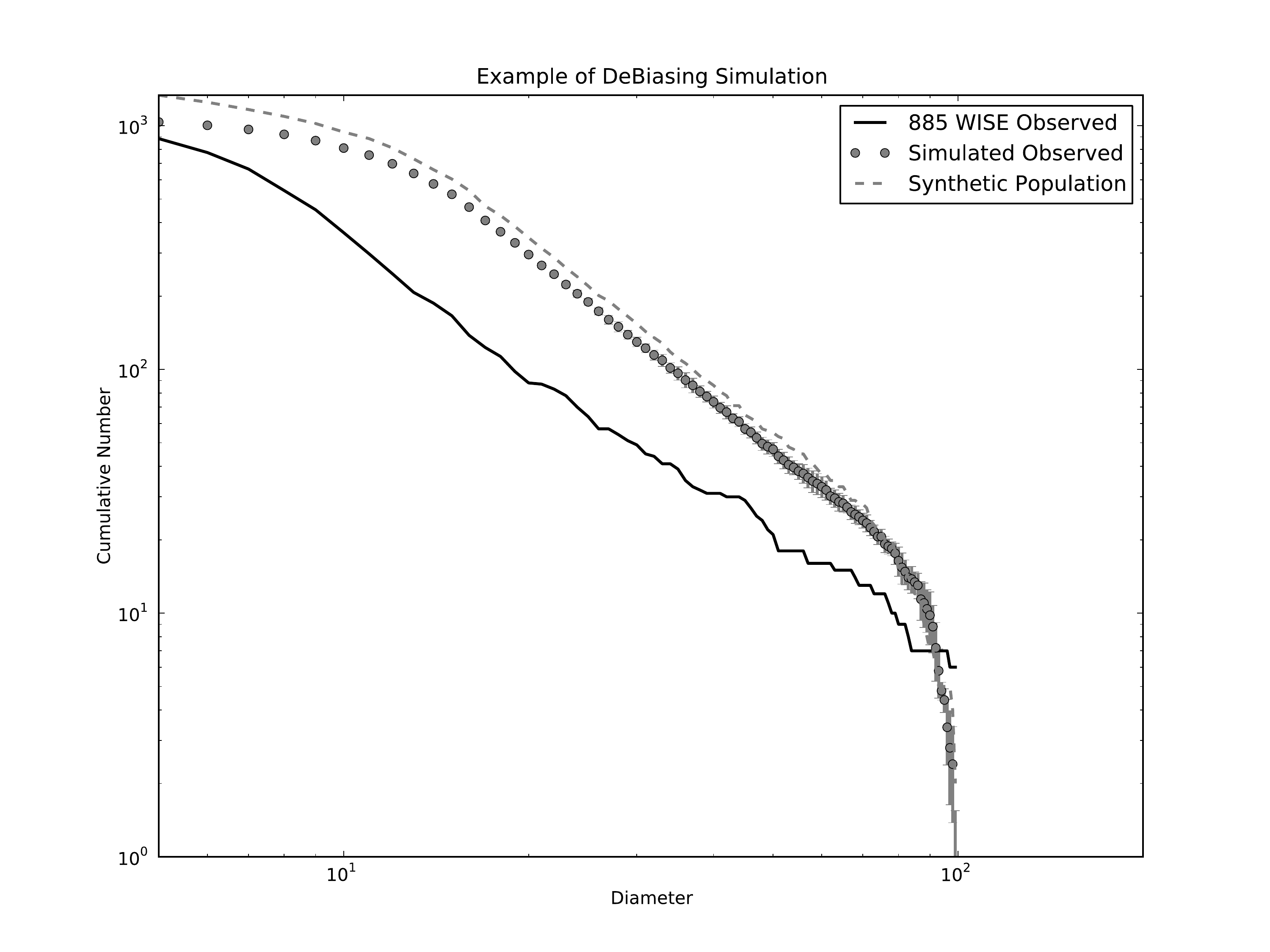}
\caption{The size distribution of the synthetic population based on the albedo, size-frequency and number distributions reported in \citet{Ryan.2011a} is shown as a dashed line (using 1334 objects larger than 5km). The resulting simulated observations, the mean and standard deviation based on 5 simulations, is shown as gray points with associated errorbars. The size-frequency of the 885 objects with $D > 5km$ observed by WISE is shown as a solid line. The resulting simulated distribution is clearly not consistent with the sample detected with NEOWISE.}
\label{fig:debias_ryan1}
\end{center}
\end{figure}

\begin{figure}[t]
\begin{center}
\includegraphics[width=12cm]{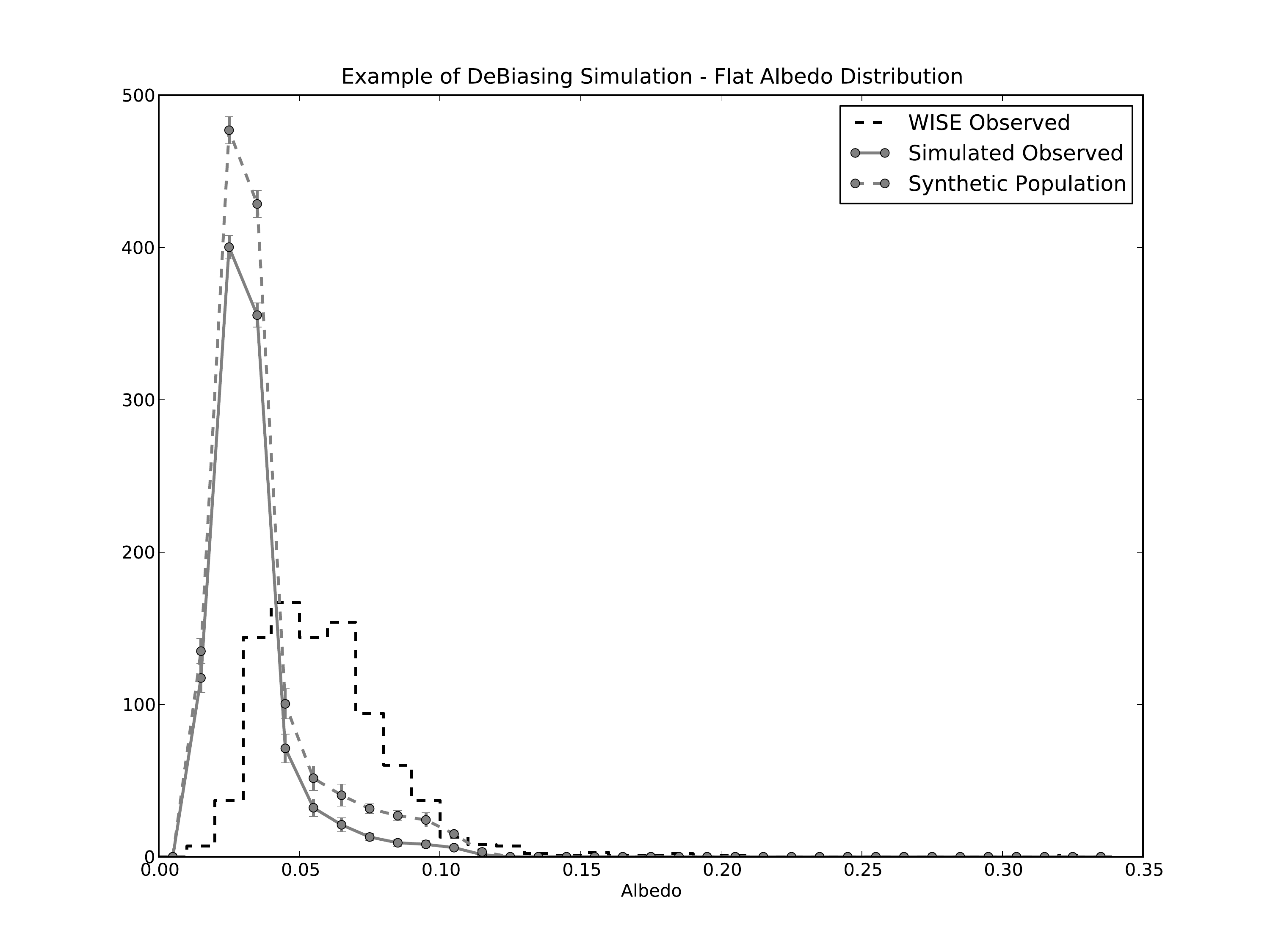}
\caption{The visible albedo, $p_V$, distributions of the synthetic population and simulated observed sample based on \citet{Ryan.2011a} are shown in gray. The resulting simulated distribution is clearly not consistent with the sample detected with NEOWISE (here shown in black). }
\label{fig:debias_ryan2}
\end{center}
\end{figure}

We then moved on to testing the albedo distribution and size-frequency derived by \citet{Ryan.2011a}. If their result is correct, we should be able to generate a synthetic population following these distributions, run this synthetic population through our survey simulator and recover a simulated observed set of objects that is nearly identical to our sample of Hildas detected by NEOWISE. The results of our simulations are shown in Figures \ref{fig:debias_ryan1} and \ref{fig:debias_ryan2}. Note that the albedo distribution of \citet{Ryan.2011a} was broadened slightly by varying the albedo of each object derived from their albedo-diameter relation by a random shift between $\pm2\%$ to account for a more realistic error estimate. The resulting simulated distributions are clearly not consistent with the sample detected by NEOWISE. 

\begin{figure}[t]
\begin{center}
\includegraphics[width=12cm]{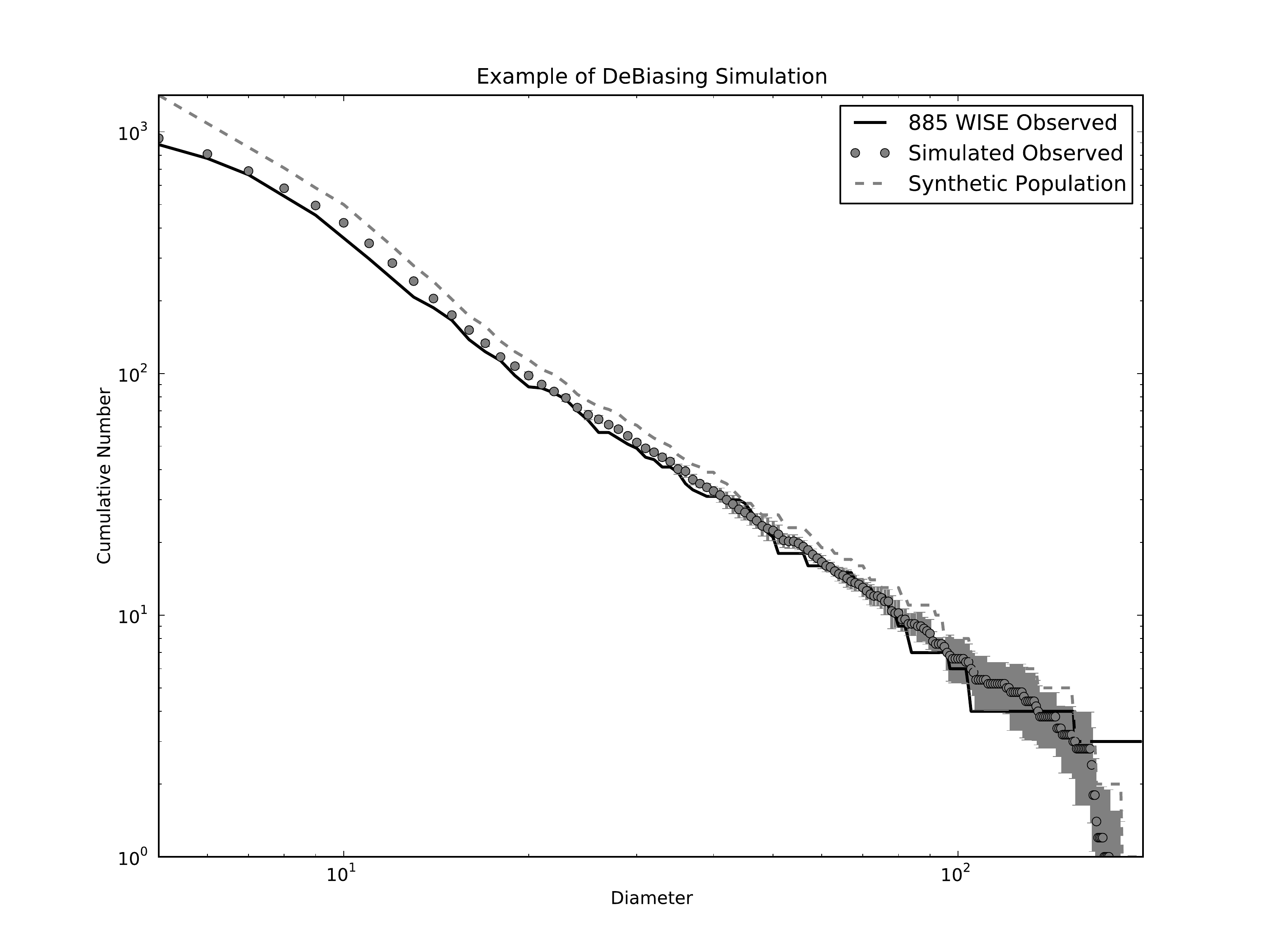}
\caption{The size distribution of the synthetic population using a single sloped power-law with slope $\alpha = 1.7$ is shown as a dashed line. The resulting simulated observations, the mean and standard deviation based on 5 simulations, are shown as gray points with associated errorbars. The size-frequency of the 885 objects with $D > 5km$ detected by NEOWISE is shown as a solid line. The resulting simulated distribution is nicely consistent with the sample detected with NEOWISE, although some refinement is clearly called for.}
\label{fig:debias2}
\end{center}
\end{figure}

\begin{figure}[t]
\begin{center}
\includegraphics[width=12cm]{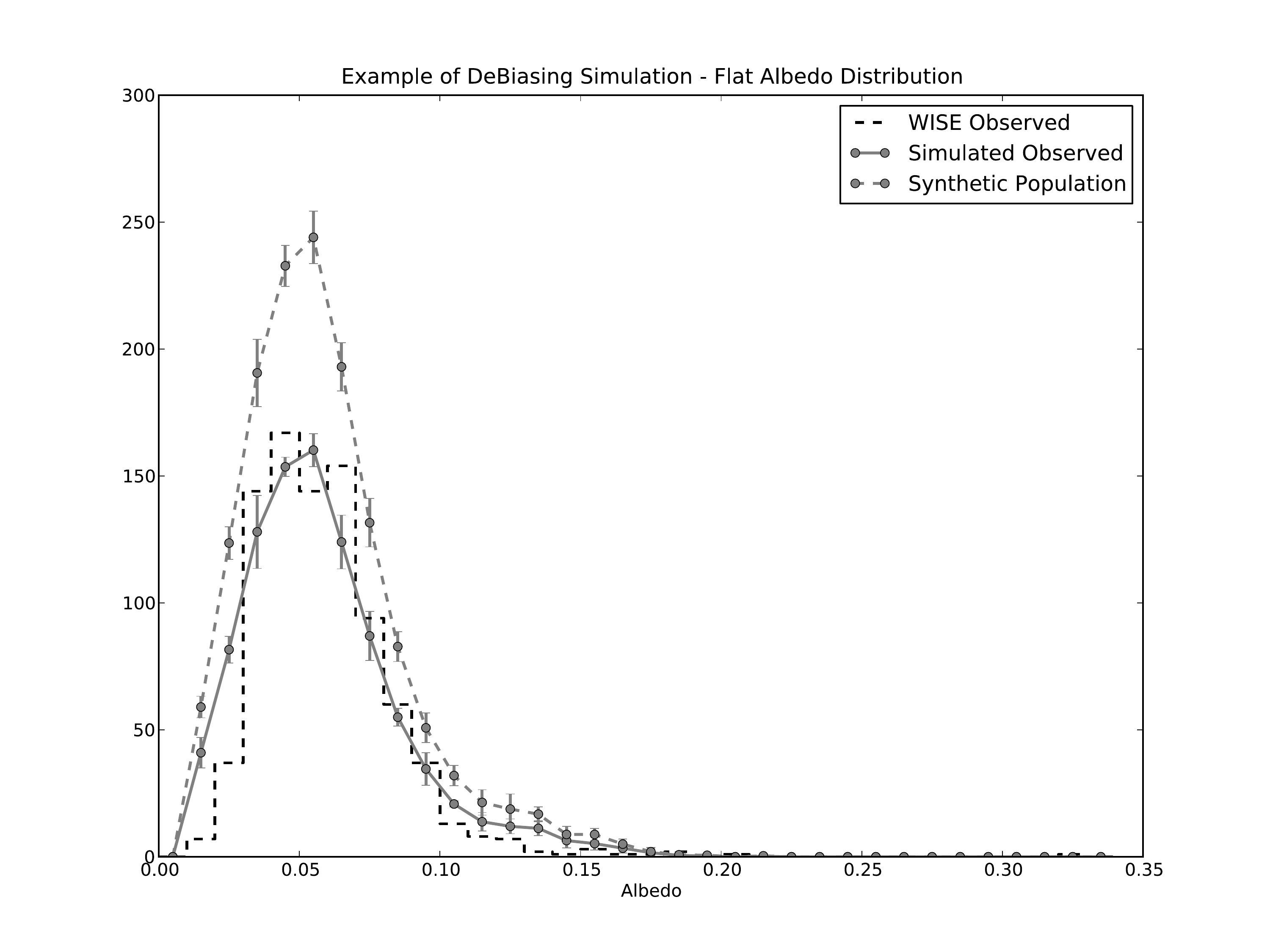}
\caption{The visible albedo, $p_V$, distributions of the synthetic population using a single Gaussian with mean and standard deviation of $0.05\pm0.03$ are shown. The resulting simulated distribution is nicely consistent with the sample observed with WISE, although some refinement is clearly called for.}
\label{fig:debias3}
\end{center}
\end{figure}

A full debiasing of the Hilda population is beyond the scope of this paper; however, we have compared the sample observed by WISE with a handful of single sloped power-laws for the size-frequency distribution and single Gaussians for the albedo distributions. An example of the resulting simulations is shown in Figures \ref{fig:debias2} and \ref{fig:debias3}. The single sloped power law with slope of $\alpha=1.7\pm0.3$ is a much better fit than the distribution given by \citet{Ryan.2011a} and no significant break at $D \sim 12$km is seen. Additional work is needed to derive refined estimates of the size-frequency and albedo distributions, and this work is underway. Future work also includes comparison of the numbers, sizes, and albedo distributions of the Hildas that we have computed to theoretical predictions based on various formation and evolution scenarios. It is, however, clear from this paper that the conclusions drawn in \citet{Ryan.2011a} are unsupported in our dataset, which consists of more than one order of magnitude additional objects in the Hilda population with thermal modeling. 

\section{Conclusions}

We derived thermal models for 1023 objects in the Hilda population, with sizes ranging from 3 to 200km, that were observed during the cryogenic part of the NEOWISE survey. The Hildas are found to have low albedo, weighted mean of $0.055\pm0.018$, with most of the objects being consistent with having a C-, P- and D-type taxonomic classification. Although there is an apparent broadening of the population at smaller sizes, this is found to be due to a natural increase in measurement outliers following Gaussian errors due to the increased number of objects at smaller sizes. For example, the weighted mean of the objects with diameter in 4-5 km range is $0.049\pm0.021$.

There are, however, a handful of objects with higher albedos and possible M- and E-type taxonomy that may be interlopers, coming from other parts of the solar system and subsequently captured in the 3:2 mean motion resonance. Furthermore, we find that the D-types dominate among the large Hildas (with $D > 30$km), making up $67^{+7}_{-15}\%$. This is compared to $26^{+17}_{-5}\%$ of these objects being C-/P-type (with the majority of these being P-type asteroids). 

We observed 219 and 112 of the members of the Hilda and Schubart collisional families, respectively. The results show that the Hilda collisional family is slightly brighter than the general population with a weighted mean albedo of $0.061\pm0.011$, while the Schubart family is significant darker with weighted mean albedo of $0.039\pm0.013$. Of the 220 objects observed in the Hilda family, 13 have derived $p_{IR}/p_V$ values that together with $p_V$ indicate that all but two are D-type asteroids. We were only able to classify 4 of the Schubart family, with all of them falling in the C-/P-type cluster in Figure \ref{fig:pVpIR}. 

We also showed that the size-frequency and size-albedo dependency found in \citet{Ryan.2011a} to be inconsistent with the results found by the NEOWISE survey. The albedos have no significant size-albedo dependency, and small objects have similarly dark surfaces as the larger objects. This results means that the size-frequency distribution derived by \citet{Ryan.2011a} is also in question. We compared their size-frequency distribution with that of a single sloped power law with $\alpha \sim 1.7\pm0.3$, and found that the latter is a much better fit to the distribution detected by NEOWISE. This suggests that the Hildas are in near-collisional equilibrium \citep{Dohnanyi.1969a} for all sizes sampled in the NEOWISE survey. More work is however needed to more accurately debias our observed sample and derive the underlaying, debiased size-frequency and albedo distributions.

\section{Acknowledgments}
This publication makes use of data products from the {\it Wide-field Infrared Survey Explorer}, which is a joint project of the University of California, Los Angeles, and the Jet Propulsion Laboratory/California Institute of Technology, funded by the National Aeronautics and Space Administration. This publication also makes use of data products from NEOWISE, which is a project of the Jet Propulsion Laboratory/California Institute of Technology, funded by the Planetary Science Division of the National Aeronautics and Space Administration. We gratefully acknowledge the extraordinary services specific to NEOWISE contributed by the International Astronomical Union's Minor Planet Center, operated by the Harvard-Smithsonian Center for Astrophysics, and the Central Bureau for Astronomical Telegrams, operated by Harvard University. We also thank the worldwide community of dedicated amateur and professional astronomers devoted to minor planet follow-up observations. This research has made use the NASA/IPAC Infrared Science Archive, which is operated by the Jet Propulsion Laboratory/California Institute of  Technology, under contract with the National Aeronautics and Space Administration.

\end{document}